\documentclass[fleqn,usenatbib]{mnras}

\usepackage[T1]{fontenc}
\usepackage{ae,aecompl}

\usepackage{graphicx}	
\usepackage{amsmath}	
\usepackage{amssymb}	

\usepackage{pifont}
\newcommand{\cmark}{\ding{51}}%

\usepackage{pgfgantt,rotating}
\usepackage{subfloat}
\usepackage{multirow}
\usepackage{blindtext}
\usepackage[flushleft]{threeparttable}
\usepackage{newtxtext,newtxmath}

\title[The MS across cosmic time]{The Main Sequence of star forming galaxies across cosmic times}

\author[P.Popesso et al.]{
P. Popesso,$^{1}$\thanks{E-mail: paola.popesso@eso.de}
A. Concas,$^{1}$
G. Cresci,$^{2}$
S. Belli,$^{3}$
G. Rodighiero,$^{4}$
H. Inami,$^{5}$
M. Dickinson,$^{6}$
\newauthor
O. Ilbert,$^{7}$
M. Pannella,$^{8}$
D. Elbaz,$^{9}$
\\
$^{1}$European Southern Observatory, Karl Schwarzschildstrasse 2, 85748, Garching bei M\"unchen, Germany\\
$^{2}$INAF-Osservatorio Astronomico di Arcetri, Largo Enrico Fermi 5, 50125, Firenze, Italy\\
$^{3}$Center for Astrophysics Harvard \& Smithsonian, 60 Garden St., Cambridge, MA 02138\\
$^{4}$Dipartimento di Fisica e Astronomia, Universit\'a di Padova, Vicolo dell' Osservatorio 3, 35122, Padova, Italy\\
$^{5}$Hiroshima Astrophysical Science Center, 1-3-2 Kagamiyama, Higashi-Hiroshima City, Hiroshima, Japan 739-8511\\
$^{6}$NOIRLab, 950 N. Cherry Ave. Tucson, AZ 85719, USA\\
$^{7}$Laboratoire d'Astrophysique de Marseille, 38 rue Frederic Joliot Curie, 13388 Marseille, France \\
$^{8}$Dipartimento di Fisica e Astronomia, Universit\'a di Trieste, Via Valerio, 2 - 34127 Trieste, Italy\\
$^{9}$Laboratoire AIM-Paris-Saclay, CEA/DRF/Irfu - CNRS - Universit\'e Paris Diderot, CEA-Saclay, 91191 Gif-sur-Yvette, France
}

\date{Accepted XXX. Received YYY; in original form ZZZ}

\pubyear{2018}

\begin{document}
\label{firstpage}
\pagerange{\pageref{firstpage}--\pageref{lastpage}}
\maketitle

\begin{abstract}
By compiling a comprehensive census of literature studies, we investigate the evolution of the Main Sequence (MS) of star-forming galaxies (SFGs) in the widest range of redshift ($0 < z < 6$) and stellar mass ($10^{8.5}-10^{11.5}$ $M_{\odot}$) ever probed. We convert all observations to a common calibration and find a remarkable consensus on the variation of the MS shape and normalization across cosmic time. The relation exhibits a curvature towards the high stellar masses at all redshifts. The best functional form is governed by two parameters: the evolution of the normalization and the turnover mass ($M_0(t)$), which both evolve as a power law of the Universe age. The turn-over mass determines the MS shape. It marginally evolves with time, making the MS slightly steeper towards $z\sim4-6$. At stellar masses below $M_0(t)$, SFGs have a constant specific SFR (sSFR), while above $M_0(t)$ the sSFR is suppressed. We find that the MS is dominated by central galaxies. This allows to turn $M_0(t)$ into the corresponding host halo mass. This evolves as the halo mass threshold between cold and hot accretion regimes, as predicted by the theory of accretion, where the central galaxy is fed or starved of cold gas supply, respectively. We, thus, argue that the progressive MS bending as a function of the Universe age is caused by the lower availability of cold gas in halos entering the hot accretion phase, in addition to black hole feedback. We also find qualitatively the same trend in the largest sample of star-forming galaxies provided by the IllustrisTNG simulation. Nevertheless, we still note large quantitative discrepancies with respect to observations, in particular at the high mass end. These can not be easily ascribed to biases or systematics in the observed SFRs and the derived MS. 
\end{abstract}

\begin{keywords}
galaxies: evolution -- galaxies: star formation -- galaxies: high-redshift
\end{keywords}



\section{Introduction}

The Main Sequence (MS) of star forming galaxies (SFGs) is considered one of the most useful tools in modern astrophysics in the field of galaxy evolution. This very tight relation between the galaxy star formation rate (SFR) and the stellar mass (M$_{\star}$) is in place from redshift $\sim$ 0 up to $\sim$ 6 \citep{2004MNRAS.351.1151B, 2007ApJS..173..267S, 2007ApJ...660L..47N,2007A&A...468...33E,2007ApJ...670..156D,2009MNRAS.393..406C,2009ApJ...698L.116P,2009A&A...504..751S,2010MNRAS.405.2279O,2010ApJ...714.1740M,2011ApJ...739L..40R,2011A&A...533A.119E,2011ApJ...730...61K,2011ApJ...738...69S,2012ApJ...754L..29W,2012ApJ...750..120Z,2012ApJ...752...66L,2012ApJ...744..154R,Salmi:2012hm,2013ApJ...767...50M,2013ApJ...777L...8K,2014MNRAS.437.3516S,2014ApJ...791L..25S,2014ApJS..214...15S,2014ApJ...795..104W,2015ApJ...804..149S,2015A&A...575A..74S,2015A&A...581A..54T,2015ApJ...801...80L,2016ApJ...820L...1K,2016MNRAS.455.2839E,2017ApJ...847...76S,2018A&A...615A.146P,2018MNRAS.477.3014B,2019MNRAS.tmp.2263P,2019MNRAS.483.3213P,2020ApJ...899...58L,Thorne2021,Leja2022,Daddi2022}. The evolution of its normalization, slope and scatter have been largely studied in the past decade. It is now well established that the normalization declines significantly but smoothly as a function of redshift, likely on mass-dependent timescales \citep[see also][]{2014ApJS..214...15S}, rather than being driven by stochastic events like major mergers and starbursts \citep{2017ApJ...844...45O}. More uncertain is the precise redshift dependence of such evolution, which is often expressed as $\propto (1+z)^{\gamma}$, with $\gamma$ varying from 1.9 to 3.7 \citep{2014ApJS..214...15S,2014ApJ...795..104W,2015A&A...575A..74S,2013A&A...556A..55I,2018A&A...615A.146P,2020ApJ...899...58L,Thorne2021,Leja2022}. This is mainly due to the uncertainty in deriving the evolution of the exact shape of the relation, which is still matter of intense debate. Several studies point to a power law shape, SFR $ \propto$ M$_{\star}^{\alpha}$, both in the local Universe \citep{2010ApJ...721..193P,2015ApJ...801L..29R} and at high redshift \citep{2014ApJS..214...15S,2014MNRAS.443...19R,2016ApJ...820L...1K,2018A&A...615A.146P}. Other works suggest that the relation exhibits a curvature towards the high mass at low \citep{2019MNRAS.483.3213P} and high redshift \citep{2014ApJ...795..104W,2015A&A...575A..74S,2015ApJ...801...80L,2015A&A...581A..54T,2016ApJ...817..118T,2019MNRAS.tmp.2263P,2020ApJ...899...58L,Thorne2021,Leja2022}. Also the scatter around the relation is quite debated with very conflicting results from the literature. Some works report a quite constant scatter of 0.2-0.3 dex from low to moderately high masses \citep[e.g.][]{2015A&A...575A..74S,2007ApJ...660L..47N,2007A&A...468...33E}. Others, instead, report a decrease of the scatter from very low ($10^8$ $M_{\odot}$) to moderate ($10^{10}$ $M_{\odot}$) stellar masses at different redshifts \citep{Willett2015,2017ApJ...847...76S,Boogaard2018}. Few others observe an increase as a function of the stellar mass from 0.3 dex at $10^{10}$ to 0.5 dex at $10^{11.5}$ $M_{\odot}$, from low to high redshift\citep[e.g.][]{2013ApJ...778...23G,2019MNRAS.483.3213P,2019MNRAS.tmp.2263P,sherman2021}.

Most of this discrepancy is ascribable to how SFGs are selected in the first place, on the SFR estimators and on the method of localization of the MS. \cite{2019MNRAS.tmp.2263P} show that color-color selection of SFGs leads to the exclusion of red dusty star-forming galaxies in particular at $z > 1.5$ and to a steep MS. \cite{2016ApJ...817..118T} show that without any selection, the MS is bending more significantly in the local than in the distant Universe \citep[see also][]{Katsianis2016}. In addition, the use of different SFR indicators might lead to systematic biases. Part of the UV emission originating from the young star population is absorbed by dust and re-processed at infrared wavelengths. Such emission alone can provide a measure of the SFR only if corrected for this absorption. However, the measure of the dust attenuation is still uncertain because of the degeneracy between age and reddening, the assumption on galaxy metallicity and SF histories, and the parametrization of the extinction curve \citep[e.g.][]{1999ApJ...521...64M,2009ApJ...703..517D,2017MNRAS.466..861D,2017MNRAS.467.1360B}. With the launch of the {\it{Spitzer}} and {\it{Herschel}} satellites in 2003 and  2009, respectively, it became possible to measure the mid and far-infrared emission for statistical samples of galaxies up to $z\sim 3$ in the most studied deep fields \citep[e.g.][]{2014ARA&A..52..373L}. Such measurements allowed finally combining the unobscured UV emission and the reprocessed far-infrared (FIR) component and calibrating them as a star formation rate indicator. Nevertheless, also such an approach might lead to systematics and uncertainty. For instance, the contamination by AGN and the overestimation of the SFR for starburst galaxies make of the mid-infrared emission, based e.g. on {\it{Spitzer}} MIPS 24 $\mu$m data, a less accurate indicator with respect to longer wavelengths \citep{2011A&A...533A.119E,Nordon:2010kh}. Nonetheless, such contamination can affect, to some extent, also the emission in the far-infrared, if the nuclear activity was particularly high in the last 100 $Myr$ of a galaxy's life. More recently, the accuracy of the SFR based on the combination of UV and IR components has been questioned by the results based on sophisticated SED fitting codes that try to reconstruct the galaxy star formation histories \citep[e.g.][]{Thorne2021,Leja2022}. Navigating through all these differences makes it very difficult to understand if the literature has reached an overall consensus on the shape and evolution of the Main Sequence across cosmic times. Furthermore, the magnitude of these effects precludes robust interpretations of derived MS properties.

To overcome this problem, and so constrain the MS evolution and systematic errors, \citet[hereafter S14]{2014ApJS..214...15S} have compiled 64 MS observations from 25 studies published in the period 2007-2014, spanning $z \sim 0$\,--\,$6$, and converted them to the same absolute calibration. These MS estimates have been taken from a variety of fields, selected using different methodologies, including both stacked and non-stacked data, and estimated with a variety of SFR indicators. By calibrating consistently all datasets, S14 determine the MS best fit as a power law, with a slope marginally evolving with redshift. The purpose of such an experiment is not to provide the "true" Main sequence. Indeed, there might still be biases and limitations related to the observational techniques, that this approach is not capable of erasing or correcting \citep[see for instance the discussion in ][]{Katsianis2020}. Rather, it aims at understanding whether the MS estimates, reported in the literature, lead to a substantial consensus once they are brought to the same calibration and all systematics are considered. It offers also the advantage that the resulting MS can be easily re-calibrated, if necessary. 

In this paper, we adopt the same approach of S14, to bring all estimates of the literature to a common calibration to check whether the community has reached or is far from reaching such a consensus. To this aim, we extend the collection of MS determination of S14 to the most recent estimates of the MS based on a variety of SFR indicators (from the combination of {\it{Spitzer}} mid-infrared and {\it{Herschel}} far-infrared data and UV emission, to SED fitting techniques) and methodology. These include additional 27 publications from 2014 to 2022, for a total of $\sim$120 MS measurements at $0 < z <6$, that we convert to the same IMF and calibrate in a consistent framework. We collect a sample of about $\sim$1500 consistently calibrated data points, that express the location of the MS as a function of stellar mass and time, and check whether there is consensus in the evolution of the MS across cosmic time.

The paper is structured as follows. Section 1 describes the MS compilation. Section 2 presents our best-fitting procedure. Section 3 shows our results. Section 4 provides a comparison of our results with previus findings and throretical predictions, while Section 5 lists our conclusions. We  assume  a  $\Lambda$CDM  cosmology  with  $\Omega_M=0.3$, $\Omega_{\Lambda}=0.7$ and $H_0=70$ \textit{km/s/Mpc}, and a Kroupa IMF throughout the paper.

\begin{table*}

\caption{ Col. 1: Reference. Col. 2: Assumed stellar initial mass function. Col. 3: Star formation rate indicator. Col. 4: redshift range analyzed Col. 5: Assumed cosmology. Col. 6: Selection methods used for the parent samples (see Appendix  or individual papers for more details). Col. 7:  type of data retrieved in the paper: {\it{stacked}} stands for stacked SFR data-points as a function of stellar mass and redshift, {\it{data-points}} for average or median SFR data-points as a function of stellar mass and redshift, and {\it{best fit}} stands for best-fit parameters expressing the MS shape as a function of stellar mass over the retrieved stellar mass and redshift ranges. Col. 8: Extinction curve reference: the details of the extinction curve implemented in MAGPHYS can be found in \protect\cite{daCunha2008}, C00 refers to \protect\cite{2000ApJ...533..682C}, CF00 to \protect\cite{2000ApJ...539..718C}, KC13 to \citet{Kriek2013} and N09 to \citet{Noll2009}. Col 9: indicates whether the data have been included in (\cmark symbol) or excluded from ($-$ symbol) the current analysis. The reasons for excluding an individual dataset from the analysis are given in Section 2. and In Appendix \ref{selectd_ms} with more details.}
\centering
\begin{tabular}{l c c c c c c c c}
    \hline
    \hline \\[-2.5mm]
Paper & IMF & SFR       & z range &$(h,\Omega_m,\Omega_\Lambda)$ & Selection & data & Extinction & included\\
      &     & indicator &         &                              &           & type & curve      & \\
      
  \hline
    \hline \\[-2.5mm]

Speagle et al. (2014)    & K   & mixed       & $0.2-6	$    & 0.7, 0.3, 0.7	& mixed\tnote{a}    	& best fit & NA & \cmark\\
Rodighiero et al. (2014) & S & IR          & $1.4-2.5$     & 0.7, 0.25, 0.75 & BzK           & stacked & NA & \cmark\\
Heinis et al. (2014)     & C & SED$+$IR    & $1.5-4$      & 0.7, 0.3, 0.7    & UV            & stacked & NA & \cmark\\
Whitaker et al. (2014)   & C & NUV$+$IR     & $0.5-2.5$    & 0.7, 0.3, 0.7   & UVJ    		& stacked & NA & \cmark\\
Chang et al. (2015)	     & C & SED$+$IR	  & $< 0.1$  	 & 0.7, 0.3, 0.7   	& mixed\tnote{b}     & best fit & MAGPHYS & \cmark \\
Lee et al. (2015)        & C & NUV$+$IR     & $0.3-1.3$	 & 0.7, 0.28, 0.72 	& NUVRJ  		& stacked & NA & \cmark\\
Ilbert et al. (2015)     & C & NUV$+$IR  	  & $0.2-1.4$ 	 & 0.7, 0.3, 0.7& NUVRJ  		& data-points & NA & \cmark\\
Tasca et al. (2015)      & C & SED         & $0.4-5$ 	 & 0.7, 0.3, 0.7   	& LBG$+I_{AB}< 25$ 	& data-points & C00 & \cmark\\
Salmon et al. (2015)   & C & SED     & $3.5--6.5$ 	 & 0.7, 0.3, 0.7 	& mixed 	& stacked & C00 & \cmark\\
Renzini \& Peng (2015)    & C & H$\alpha$   & $<0.085$ 	 & 0.7, 0.3, 0.7   	& mixed\tnote{c}     & best fit & CF00 & \cmark \\
Schreiber et al. (2015)  & S & FUV$+$IR     & $0.3-4$ 	 & 0.7, 0.3, 0.7   	& UVJ 			& stacked & NA & \cmark\\
De los Reyes et al. (2015) & S &  H$\alpha$    & $0.8$ 	 & 0.7, 0.3, 0.7   	& H$\alpha$ emission 	& stacked & CF00 & \cmark\\
Erfanianfar et al. (2016)& C & IR          & $0.2-1.5$ 	 & 0.7, 0.3, 0.7   	& mixed\tnote{b}     & data-points  & NA & \cmark\\
Tomczak et al. (2016)    & C & NUV$+$IR     & $0.2-3 $	 & 0.7, 0.3, 0.7   	& UVJ 			& stacked & NA & \cmark\\
Santini et al. (2017)    & S & SED      	  & $1.3-6 $	 & 0.7, 0.3, 0.7   	& mixed 	& best fit & C00 & \cmark\\
Kurczynski et al. (2016) & S & SED         & $0.5-3 $	 & 0.7, 0.3, 0.7   	& UVJ 			& best fit & C00 & \cmark\\
Pearson et al. (2018)    & C & SED$+$IR    & $0.2-6 $	 & 0.704, 0.272, 0.728 	& UVJ 		& best fit & CF00 & --\\
Belfiore et al. (2018)   & C & H$\alpha$   & $<0.1 $	     & 0.7, 0.3, 0.7   	& mixed\tnote{d}     & best fit & CF00 & \cmark\\
Davidzon et al. (2018)   & C & GSMF        & $2-6$       & 0.7, 0.3, 0.7         & NUVRJ     & stacked & C00 & \cmark\\ 
Lee et al. (2018)  & C & FUV$+$IR     & $1.2-4 $	 & 0.7, 0.3, 0.7   	& UVJ 			& stacked & NA & \cmark\\
Iyer et al. (2018) & C & SED    & $0.5-6 $	 & 0.7, 0.3, 0.7   	& mixed 	& stacked & C00 & \cmark\\
Popesso et al. (2019a)   & C & IR/H$\alpha$& $<0.085 $	 & 0.7, 0.3, 0.7 	& mixed\tnote{c}	& data-points & NA & \cmark\\
Popesso et al. (2019b)   & C & FUV$+$IR     & $0.2--2.5$ 	 & 0.7, 0.3, 0.7 	& mixed\tnote{c} 	& data-points & NA & \cmark\\
Barro et al. (2019)   & C & SED     & $0.5--3.$ 	 & 0.7, 0.3, 0.7 	& UVJ 	& best fit & C00 & \cmark\\
Leslie et al. (2020)   & C & radio     & $0.3--6$ 	 & 0.7, 0.3, 0.7 	& NUVRJ 	& stacked & NA & \cmark\\
Thorne et al. (2021)   & C & SED     & $0--9$ 	 & 0.678, 0.308, 0.692 	& sSFR cut 	& best fit & CF00  & \cmark\\
Sherman et al. (2021) & C & SED     & $1.5--3$ 	 & 0.7, 0.3, 0.7 	& mixed 	& best fit & KC13 & \cmark\\
Leja et al. (2022)   & C & SED     & $0.2--3.$ 	 & 0.698, 0.235, 0.765 	& ridged line 	& best fit & N09 & \cmark\\
  \hline
    \hline \\[-2.5mm]
\end{tabular}

\label{tt}
\end{table*}

\section{The MS collection}
\label{criteria}

In this work, we focus on MS estimates that have been published after 2014. To consider in the analysis the MS measurements that have been published in the period 2007-2014, we include here the MS relation determined by S14 as a result of a compilation of 64 MS estimates collected in 25 papers (see Table 3 of S14). We consider, in particular, the fit n. 64, which S14 provide as their reference MS.  
To this, we add all the MS relations that satisfy the following criteria: 
\begin{enumerate}
\item Includes a published M$_{\star}$\,--\,SFR or M$_{\star}$\,--\,sSFR (sSFR=SFR/M$_{\star}$) relation (slope $\alpha$ and normalization $\beta$) or otherwise analogous quantities;
\item Fit(s) includes more than two data points (if stacked) or 50 galaxies (if directly observed). This is required to avoid biases resulting from small number statistics;
\item Stacked points must provide mean or median of more than 25 points, to avoid large uncertainties due to low number statistics; 
\item Published after 2014.
\end{enumerate}

The 27 publications that are retrieved considering these criteria, in addition to S14, are listed in Table \ref{tt}, together with the used IMF, SFR indicator, redshift range, cosmological parameters, SFGs selection method, and extinction curve.  In Appendix, we provide a full description of the individual publication data and of their calibration.

All the considered MS estimates are based on the deepest UV, optical, IR and radio surveys ever realized on the CANDELS, COSMOS and ECDFS fields at intermediate and high redshift, and on the WISE, GALEX and optical spectroscopy in the SDSS area at $z\sim$0. Of these, 10 publications \citep{2014MNRAS.437.1268H, 2014ApJ...795..104W, 2015ApJS..219....8C,2015ApJ...801...80L, 2015A&A...579A...2I, 2015A&A...575A..74S, 2016ApJ...817..118T, 2018A&A...615A.146P, 2018ApJ...853..131L, 2019MNRAS.tmp.2263P} out of 27 are based on an SFR indicator given by the combination of UV and IR data. Additional 3 are based only on far-infrared PACS data \citep{2014MNRAS.443...19R,2016MNRAS.455.2839E, 2019MNRAS.tmp.2263P}. Other 3 \citep{2015ApJ...801L..29R, 2018MNRAS.477.3014B} are based on $H\alpha$ derived SFR taken from the SDSS spectroscopic dataset \citep{2004MNRAS.351.1151B} or from 3D-HST data at higher redshift \citep{delosreyes2015}. Additional 9 include SFR derived through SED fitting technique based on different fitting methods and star formation history reconstructions \citep{2015A&A...581A..54T,Salmon2015,2017ApJ...847...76S, 2016ApJ...820L...1K,Iyer2018,barro2019,Thorne2021,sherman2021,Leja2022}. Of the rest, one is based on the evolution of the galaxy stellar mass function of star-forming galaxies \citep{2018ApJ...852..107D}, and one is based on deep radio data \citep{2020ApJ...899...58L}. S14 is based on a collection of different SFR estimators. According to the definition of S14, all of the considered publications but two are based on "mixed" methods (see next section) for the selection of SFGs. These include color-color techniques (BzK, UVJ, and NUVRJ), 2$\sigma$ clipping, and bimodality in the SFR-M$_{\star}$ plane. Only 3 publications have an MS obtained for blue and active galaxies selected in the UV or with the Lyman Break Galaxy (LBG) technique.   

Differently from S14, we include in our study also the MS estimate at $z\sim0$. S14 discuss the inability to distinguish a "best" MS fit among the available $z\sim0$ estimates obtained before 2014. Thus, they decide not to include the local MS estimates. \cite{2019MNRAS.483.3213P} discuss extensively all the datasets available in the local Universe, selection effects and different SFR estimators and the level of agreement between the different estimates. On the basis of these results, we conclude that the inter-publication scatter of the local MS estimates available in the period 2014-2022 is comparable to that found at higher redshifts. Indeed, no different level of agreement is found as a function of redshift, once all MS estimates are brought to the same calibration (see next section for more details). 

From each publication either we take the mean or median SFR data points or staked SFR data points at the observed stellar mass and redshift without any extrapolation or interpolation, or, if these are not available, we use the provided MS best-fit parameters at a given redshift to estimate the MS in the provided stellar mass range in bins of 0.15 dex in stellar mass. The bin width is chosen to be representative of the average stellar mass error of the considered papers. The mass ranges have either been taken directly from the paper in question or estimated based on the data included in the relevant fits, rounded to the nearest $0.1$\,dex (see Appendix for a detailed description of the data taken from each publication). This leads to $\sim$120 determinations of the MS for a sample of $\sim$ 1500 data points of MS SFR as a function of stellar mass and redshift or time ($SFR(M_*,z)$ or $SFR(M_*,t)$). The data collected here encompass the widest range in redshift ($0 < z < 6$), stellar mass ($10^{8.5}-10^{11.5}$ $M_{\odot}$), and SFR ($0.01-500$ $M_{\odot}$yr$^{-1}$) available in the literature, and intend to give a census of most of the techniques and methods used to derived the MS location. Following the example of S14, we present what we hope is the broadest and most accurate census of MS observations to date.

\subsection{MS calibration}

As underlined in S14, several aspects need to be taken care of when comparing different MS estimates and before attempting any analysis:  

\begin{enumerate}
\item Initial Mass Function;
\item SFR estimator;
\item SPS model;
\item cosmology;
\item emission line effect in the estimate of SFR and M$_{\star}$;
\item Star formation histories (SFH);
\item dust extinction curve;
\item photo-z biases;
\item SED fitting procedures;
\item selection effects due to different SFG population selection.
\end{enumerate}

A detailed discussion in S14 points out that only the points {\it{i)}}, {\it{ii)}} and {\it{iii)}} lead to relevant corrections when calibrating all the MS estimates to a common ground. Instead, different cosmologies ({\it{iv}}) have relatively negligible effects ($< 0.05$\,dex) at $0 < z < 6$, which is the same redshift range considered here. The effect of emission lines ({\it{v}}) in the estimate of M$_{\star}$ and SFR is relevant only when the multi-wavelength information are limited to few photometric bands. They are, instead, negligible for datasets like CANDELS, and COSMOS, as those considered here. 

Similarly to S14, we choose not to adjust our results for differences in assumed star formation history ({\it{vi}}), dust attenuation curves ({\it{vii}}), possible photo-z biases ({\it{viii}}), or differences in SED fitting procedures ({\it{ix}}). This might differ substantially \citep[see the extensive discussion in][]{Simha2014,Acquaviva2015,Salmon2015,Ciesla2017,Iyer2017,Theios2019,Carnall2019,Lower2020,Thorne2021,Curtis-Lake2021}. Testing and correcting the effect due to such differences would require checking the galaxy SFR catalogs used to build the MS of the individual publications. In none of the considered cases, such information is publicly available. Only the stacked, averaged, or fitted MS is provided, which prevents performing any correction. In principle, the effect of the different assumptions and implementations could be predictable. There is a large recent literature of such attempts based on the combination of dust radiative transfer models and simulations \citep{Baes2020,Trayford2020,Lower2020,Narayanan2021,Lovell2021,Katsianis2021}. However, such corrections might introduce biases dependent on the different assumptions of the models. Thus, we choose not to apply such corrections. As shown at the end of the next section, neglecting the possible effects of the mentioned differences does not lead to an increase of the resulting inter-publication scatter.

We apply a correction only for 2 publications \citep{Thorne2021, Leja2022}, for which the authors explicitly indicate a systematic difference in the estimate of the stellar mass with respect to previous works of our collection. This correction is applied not because we consider these quantities wrong or inaccurate, but only to bring all MS estimates to the same stellar mass and SFR scale (see next paragraph and Appendix \ref{selected_ms}). All other publications based on the SED fitting technique, do not report systematic differences in their stellar mass and SFR estimates. Thus, no further corrections are applied.

SFG selection methods ({\it{x}}) can lead to substantially different MS slopes. S14 distinguish between "bluer", "mixed", and "non-selective" techniques: "bluer" methods are based on a simple color cut to select blue galaxies, "mixed" ones involve a more sophisticated distinction between star-forming and quiescent galaxies, while "non-selective" ones do not apply any selection. In particular, S14 find that "bluer"-based ("non-selective"-based) MS slopes are biased towards values closer to unity (zero), with respect to "mixed"-based slopes. 

To bring every relation onto a common framework, we apply the following steps. 
We use the Equations reported in Section \ref{ms_corr} to convert the M$_{\star}$ values to a Kroupa IMF, and the SFR estimates to the \citet[hereafter KE12]{kennicutt_evans_12} calibration (based on the Kroupa IMF). The choice of the KE12 calibration is dictated by the fact that it is the only one available in the literature that calibrates consistently all the SFR indicators considered here, with the exclusion of the SED fitting technique. Thus, it is the obvious choice to bring most of the SFR estimates to a common framework. The available SFRs based on the SED fitting technique included in the collection, are compared case by case with the KE12 calibration to check for consistency, as described in Appendix \ref{selected_ms}. The KE12 calibration is based on data of star forming regions in a nearby galaxy \citep{Murphy2011,Hao2011}. To check that the SFR indicators calibrated according to KE12 provide consistent estimates also at higher redshift, we perform several tests in Appendix \ref{A2}. The result of such tests is that SFRs derived through $IR$, $UV+IR$, $H{\alpha}$ and radio emission at 1.4 GHz according to the KE12 calibration (see next section), are all consistent up to $z\sim 3$ with a scatter that varies from 0.2 to 0.25 dex.

We adjust for differences in cosmology using the $(h,\Omega_M,\Omega_\Lambda)=(0.7,0.3,0.7)$ WMAP concordance cosmology \citep{spergel+03}. Furthermore, no additional dust correction is required, as the dust has been corrected for in the considered publications, and different extinction curves do not have a significant impact on the MS determination (see S14). Finally, as only 3 \citep{2015A&A...581A..54T,2014MNRAS.437.1268H,Thorne2021} out of the 27 MS estimates considered here are consistent with a "bluer" selection, we do not correct in this respect, but we discuss possible biases on the result.

As pointed out in \cite{2019MNRAS.483.3213P}, another source of small discrepancy among different MS estimates is the method used to determine the MS location, e.g. as the mean or median SFR of the SFG population in the MS region. A correction can be made in this respect under the assumption that the SFR distribution at fixed M$_{\star}$ is log-normal in the MS region. This assumption is justified by several works in literature, which find a log-normal distribution at any stellar mass \citep{2007ApJ...670..156D,2011ApJ...739L..40R,2015A&A...575A..74S,2019MNRAS.483.3213P,2019MNRAS.tmp.2263P, Leja2022}. In this case, the peak of the distribution coincides with the median SFR. The mean SFR is always larger than the median by an offset that depends only on the dispersion of the distribution. For a dispersion of $\sim0.3$ dex, the correction is $\sim0.1$ dex and it increases to $\sim0.17$ dex for a dispersion of $\sim0.5$ dex. As pointed out in the Introduction, many discrepant results in the literature do not allow us to reach a consensus on the value of the MS scatter as a function of the stellar mass. Thus, we assume an average dispersion of 0.3 dex and apply the corresponding correction in the further analysis to all MS estimates based on the median SFR to convert them into mean values. 

\begin{figure*}
\includegraphics[width=\columnwidth]{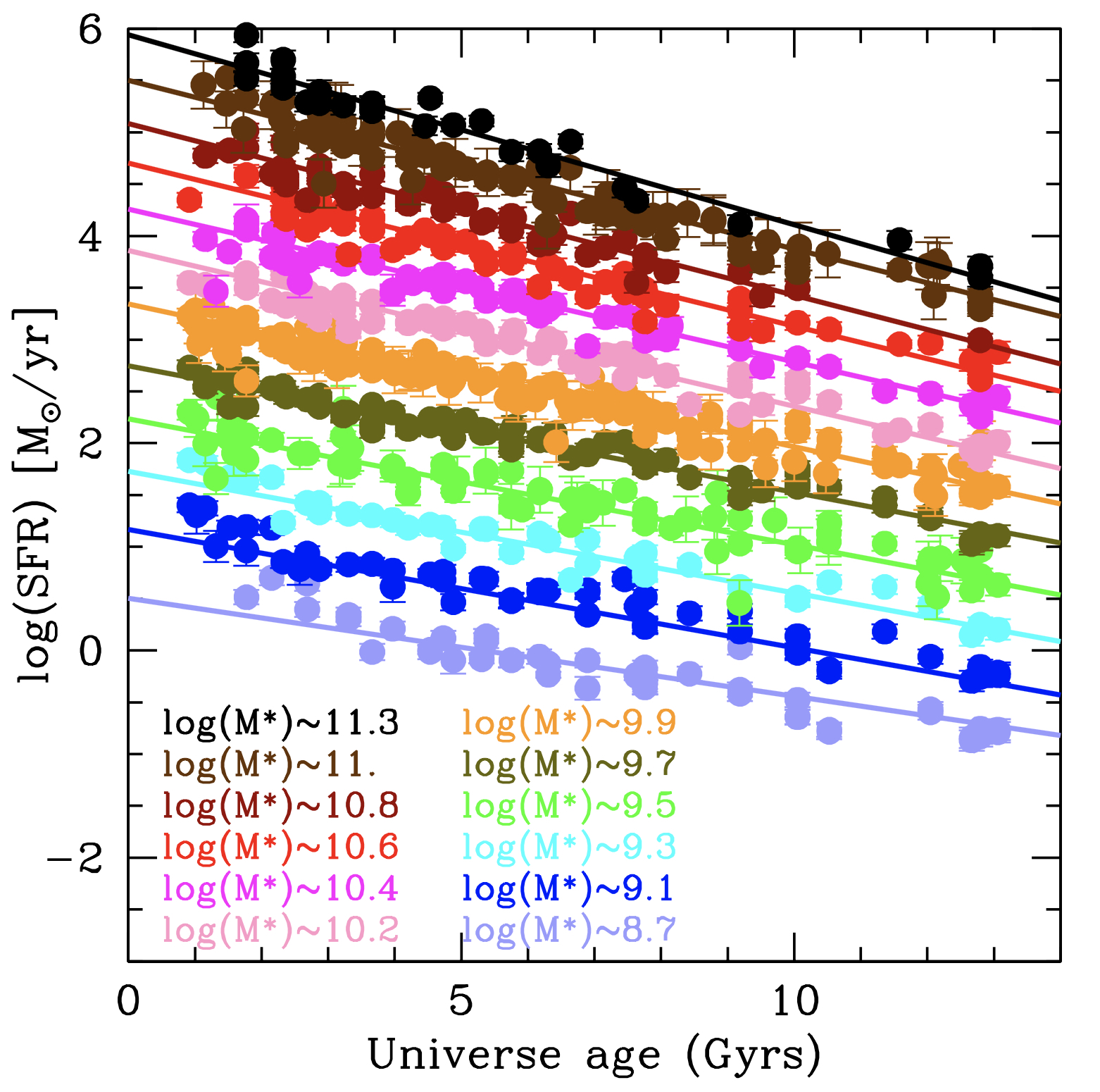}
\includegraphics[width=\columnwidth]{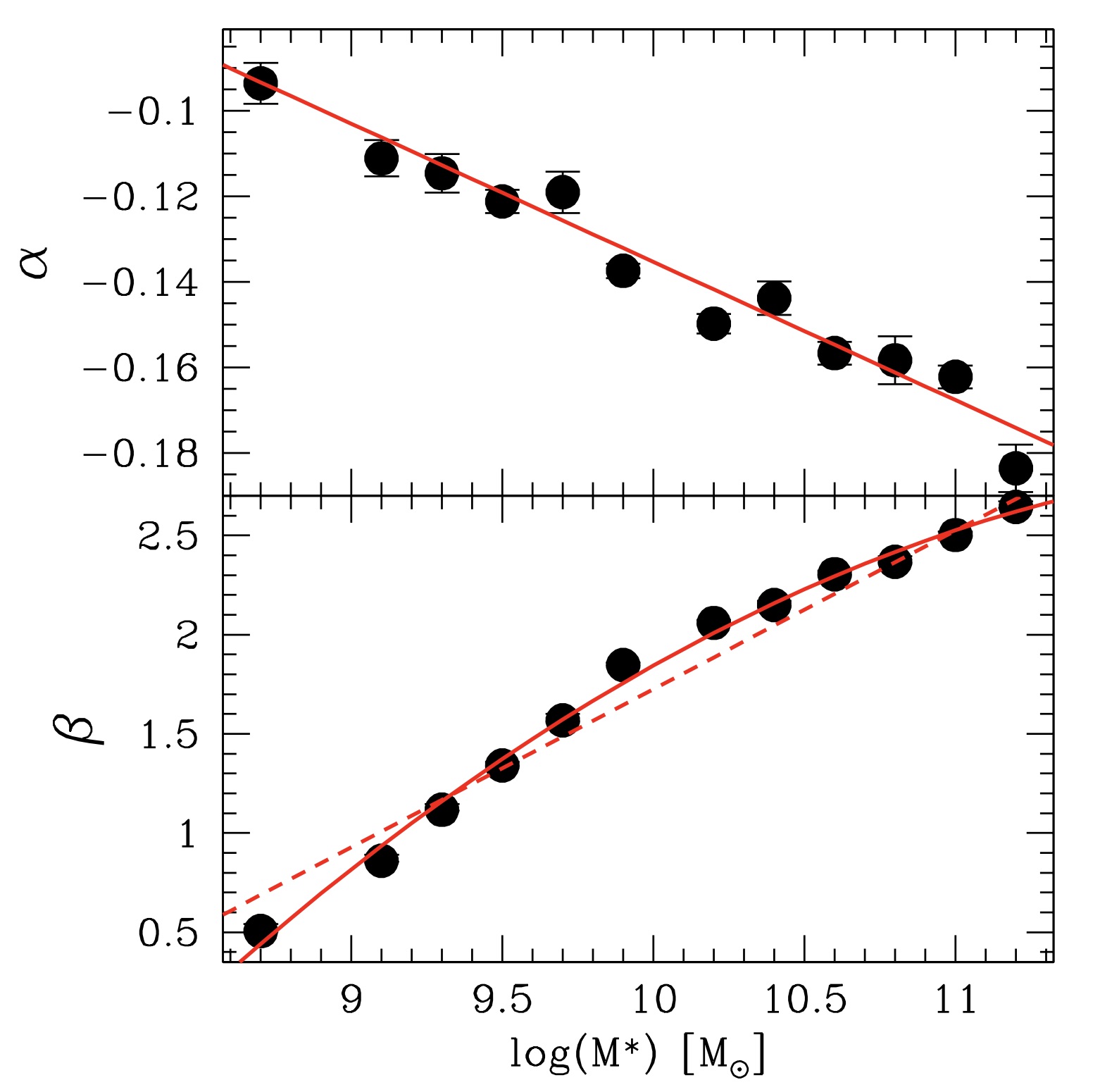}
\caption{{\it{Left panel:}} logSFR versus Universe age in several bins of stellar masses. The data points indicate the SFR based on the MS estimates collected in this work. The solid lines show the best linear fit as in Eq. \ref{ciccio}. Data points and lines are color-coded as a function of the stellar mass bin as indicated in the figure. For clarity, the relations are artificially displaced by 0.4 dex from one another. {\it{Right panel:}} $\alpha(logM_{\star})$ (upper panel) and $\beta(logM_{\star})$ (lower panel) as a function of $logM_{\star}$. The red solid lines in both panels indicate the best fit relations of Eq. \ref{ciccio}.  The dashed line in the bottom panel shows the liner fit approximation as proposed in S14.}
\label{linear}
\end{figure*}

\subsection{The main corrections}
\label{ms_corr}
We describe here the corrections applied to bring the different MS estimates to a common framework:

\begin{itemize}
\item {\bf{IMF correction}}\\
We apply the IMF offsets to stellar masses as in S14 with the form:
\begin{equation}\label{eq:imf}
M_{\star,K} = 1.06\,M_{\star,C} = 0.62\,M_{\star,S},
\end{equation}
with the subscripts referring to Kroupa, Chabrier, and Salpeter IMFs, respectively. These correspond to stellar mass offsets of $0.03$ and $-0.21$\,dex, for the Chabrier and the Salpeter IMF, respectively. 
    
\item {\bf{SFR indicator correction}}\\
All SFR estimates have been converted to the calibration of KE12, which is based on the Kroupa IMF. The KE12 prescriptions are different depending on whether $\nu{L_{\nu}}$ is estimated in the FUV ($\sim$1300$-$1700\AA) or NUV ($\sim$2300$-$2800\AA). Following Tables 1 and 2 of KE12, we make use of the following calibrations:

\begin{equation}
    SFR(FUV+IR)=1.71\cdot10^{-10}(L_{FUV}+0.46*L_{IR})
    \label{FUV_IR}
\end{equation}
\begin{equation}
    SFR(NUV+IR)=2.06\cdot10^{-10}(L_{NUV}+0.27*L_{IR})
    \label{NUV_IR}
\end{equation}
\begin{equation}
    SFR(H\alpha)=2.06\cdot10^{-8}(L_{H\alpha})
    \label{Ha}
\end{equation}
\begin{equation}
    SFR(IR)=1.49\cdot10^{-10}(L_{IR})
    \label{IR}
\end{equation}
\begin{equation}
    SFR(1.4GHz)=2.43\cdot10^{5}(L_{1.4GHz})
    \label{radio_sfr}
\end{equation}
where $L_{IR}$, $L_{FUV}$, $L_{NUV}$ and $L_{H\alpha}$ are the luminosities (in solar luminosity units) estimated in the IR (range $8-1000$ $\mu$m), FUV and NUV, $H\alpha$ emission line and radio emission (1.4 GHz), respectively.
The SFRs based on SED modeling, instead, are corrected only for the IMF, according to the derivative of Eq. \ref{eq:imf}. The correction to a common SFR indicator varies from 0.05 to $-0.2$ dex. After applying the calibration, the local MS of \cite{2015ApJS..219....8C} is still systematically lower with respect to the other local MS estimates by 0.15 dex (see \citealt{2019MNRAS.483.3213P} for a detailed comparison with UV$+$IR based and H$\alpha$ based SFR). Thus, we correct for this offset as indicated in \cite{2019MNRAS.483.3213P} before including the MS estimate of \cite{2015ApJS..219....8C} into our sample.

 As already pointed out in \cite{2019MNRAS.483.3213P}, \cite{2018A&A...615A.146P} report a systematic offset of their SFR of $\sim$0.4 dex below all other SFR estimates considered here at the same redshift. As shown in \cite{2019MNRAS.483.3213P}, their MS lies below all other determinations at more than 1$\sigma$ at all redshift. \cite{2011A&A...533A.119E} show that SPIRE and PACS SFR estimators lead to consistent results. So we conclude that the discrepancy must be related to the deblending technique of the SPIRE detections and the SED fitting technique applied in \citet[][see their Appendix C for an extensive discussion]{2018A&A...615A.146P}. Since the problem might be related to an over-deblending issue rather than to the SFR indicator, we decide not to correct the SFR of \cite{2018A&A...615A.146P} and to exclude those MS estimates from the dataset. 
 
In addition, two of the considered publications \citep{Thorne2021, Leja2022} based on two different SED fitting algorithms, {\it{ProSpect}} \citep{Robotham2020}, and {\it{Prospector}} \citep{Leja2017}, report lower values of SFR and larger values of stellar masses with respect to the rest of the MS estimates considered here. This discrepancy is due to a different reconstruction of the galaxy star formation history with respect to other SED fitting methods. The main difference consists in the fact that both codes include the contribution of an extra component of stars older than 100 $Myr$, which would affect both the stellar mass and the SFR estimates. The effect on the stellar masses is obvious and it is reported to be 0.2 dex for \citet{Thorne2021} and 0.3 dex for \citet{Leja2022}. \citet{Leja2022}, in particular, report also that the derived SFRs tend to be lower at fixed stellar mass with respect to $IR$ or $UV+IR$ derived SFR in the redshift range $0.5 < z < 3$. They ascribe such a difference to the contribution of the extra component of old stars to the dust heating, which would increase the galaxy IR emission. They also find that such contribution is redshift dependent, being larger at $0.5 < z < 3$ than at lower and higher redshifts. \citet{Leja2022} conclude that $IR$ or $UV+IR$ derived SFRs are overestimated at $0.5 < z < 3$, because not all the IR emission is due to dust heating by young stars and, thus, to star formation. 

The KE12 calibration does take into account that only a fraction of the $IR$ emission is due to star formation. However, it assumes that such a fraction is the same at all redshifts. If this assumption is not valid, it implies that at higher redshift, the $IR$ or $IR+UV$ SFR indicators based on the KE12 calibration might lead to over or underestimated SFRs with respect to the radio or H$\alpha$ derived SFRs, based on the same calibration. This is because the contribution of an extra old star component might significantly affect the IR emission, but it has rather a negligible effect on the H$\alpha$ and radio emission, which are dominated by HII region nebular emission and non-thermal (synchrotron) emission due to Type II and Type Ia Supernovae, respectively. As shown in the Appendix \ref{A2}, we do not observe such discrepancy at any redshift. Instead, the KE12 calibrated SFR derived through the radio, H$\alpha$ and $IR+UV$ emission are consistent from $z\sim 0$ up to $z\sim 3$. Thus, we do not find clear evidence of an evolution of the fraction of $IR$ emission contributing to the SFR. In addition, we point out that \citet{Thorne2021} show that once the stellar mass discrepancy is taken into account, the SFRs derived with {\it{ProSpect}} are perfectly consistent with the radio emission based SFRs of \citet{2020ApJ...899...58L}. We perform the same exercise between the {\it{Prospector}} MS estimates of \citet{Leja2022} and those of \citet{2014ApJ...795..104W} based on $IR+UV$ emission on the same 3DHST galaxy sample in the CANDELS fields (see Fig. \ref{leja_whit} in Appendix \ref{selected_ms}). Once the stellar masses are corrected for the discrepancy (0.3 dex), the MS estimates are consistent within 1-1.5$\sigma$ at all stellar masses. Thus, we include the \citet{Thorne2021} and \citet{Leja2022} MS estimates by correcting the stellar masses by the reported discrepancy, and we leave unaltered the SFR estimates.

\item {\bf{Cosmology correction}}: this is calculated as the ratios between luminosity distance, $d_L(z)$, derived from two different cosmologies, and, given the observed redshift range of a sample, applying a $d_L^2$ correction at the expected median $z$ of galaxies in the sample. S14 estimate also first-order volume effects. However, they find that this account for a negligible effect in all cases. Thus, we do not take it into account.

\end{itemize}

After applying the calibration, we obtain an inter-publication scatter of 0.08 dex per bin of time and stellar mass. We observe also that the scatter is not much dependent on the SFR indicator or methodology used in the different publications. If we limit the analysis only to publications based on $IR$ or $UV+IR$ based SFRs or only to those based on the SED fitting technique, we do not observe any scatter variation.

\subsection{Selection effects}
\label{selection_effect}

While efficient at selecting SFGs, most selection techniques differ from each other and do not all select the same population. S14 discuss extensively that the selection of blue objects preferentially select actively star-forming, non-dusty galaxies and exclude a large percentage of galaxies that are classified as SFGs via other selection mechanisms (e.g. color-color selection). This leads on average to larger MS slopes than the ones retrieved with other selection methods. One of the methods to retrieve blue sources is the Lyman break technique (\citealt{steidel+99}; \citealt{stark+09}; \citealt{bouwens+11}), used to select high-$z$ Lyman-break galaxies (LBGs). In the list of publications considered here only \cite{2015A&A...581A..54T} apply the LBGs selection technique. \cite{2016ApJ...817..118T} compare the results of \cite{2015A&A...581A..54T} with those based on the "mixed" selection method (see also discussion below) and find very good agreement. We conclude that the results of \cite{2015A&A...581A..54T} seem less affected by the bias of the bluer selection discussed by S14. Similarly, \cite{2014MNRAS.437.1268H} apply a UV selection to identify distant SFGs. Nevertheless, their estimates are not scattering significantly with respect to the other relations. Thus, we conclude that also in this case, the UV selection does not affect significantly the slope of the relation. \citet{Thorne2021} applies a sSFR cut, which might similarly select blue galaxies. However, the authors report a very good agreement with the MS estimates of \citet{2020ApJ...899...58L}, which is based on the radio selection. Thus, we conclude that also in this case the bias is negligible.

All other MS estimates included in our analysis are based on methods that S14 classify as "mixed". Such techniques include redder objects in the selection, thus considering also a large portion of the SFG population dominated by dust. These methods are considered to provide a more physical distinction between SFGs and quiescent galaxies \citep{2013A&A...556A..55I,2015A&A...575A..74S}. Among these "mixed" methods, we include the color-color selection based on the rest-frame (U-V) $-$ (V-J) and (M$_{NUV}$-M$_R$) $-$ (M$_R$-M$_J$) absolute colors, the 2$\sigma$ clipping, the bimodality between SFGs and quiescent galaxies in the SFR-M$_{\star}$ plane. While these different selection methods do not seem to affect the average observed SFRs across different publications, as pointed out by S14, they do seem to influence the derived slopes and the intrinsic scatter of the MS. \cite{2019MNRAS.tmp.2263P} show that all these methods agree very well over most of the stellar mass range considered. However, small discrepancies can be observed due to little selection biases at very high stellar masses towards high redshift. Namely, selections as the (U-V) $-$ (V-J) color selection tend to exclude part of the high mass star-forming galaxies at relatively lower SFR with respect to the MS location. This would lead to a steeper MS at higher redshift with respect to the low redshift relation and thus a more significant evolution. The effect is an offset of $\sim$0.15$-$0.2 dex at M$_{\star}$ of $10^{11}-10^{11.5}$ $M_{\odot}$. Instead, the selection based on the (M$_{NUV}$-M$_R$) $-$ (M$_R$-M$_J$) absolute colors is less prone to this selection bias, as it allows to select all galaxies populating the log-normal distribution around the MS location \citep{2019MNRAS.tmp.2263P,2015A&A...579A...2I}. 

We point out that the MS estimates affected by this bias are the stacked points based on the CANDELS field dataset \citep{2014ApJ...795..104W, 2015A&A...575A..74S, 2016ApJ...817..118T}. Due to the very small volume sampled by the CANDELS fields, these include less than 15 galaxies per stacked point at stellar masses larger than $10^{11}$ $M_{\odot}$. Since this is below the limit required in Section \ref{criteria}, those stacked points are anyhow not included in our analysis. 

\begin{figure}
\includegraphics[width=\columnwidth]{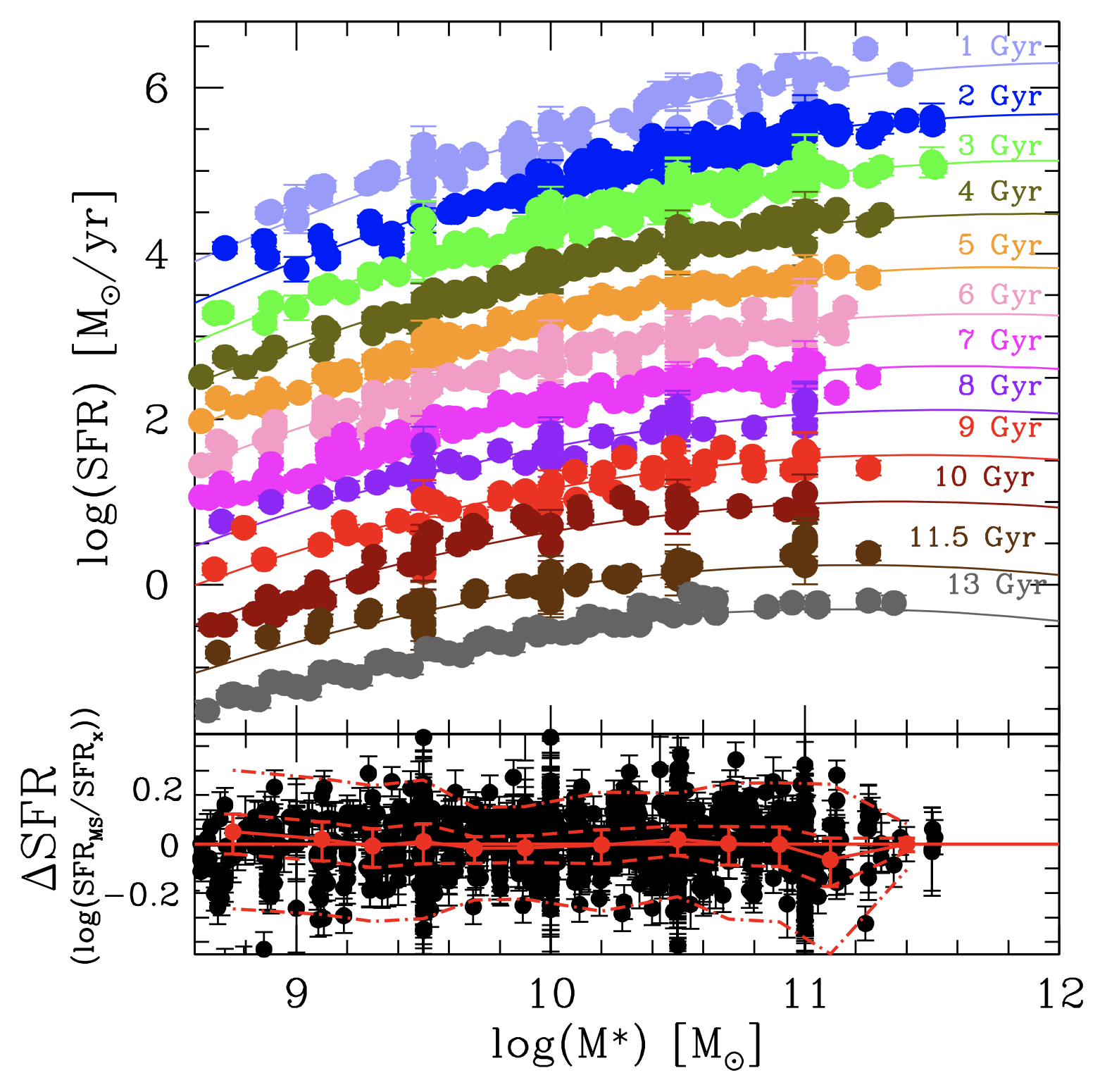}
\caption{The upper panel shows the MS shape as a function of the Universe age from 1 to 13 Gyrs. The data points show the MS estimates collected and calibrated in this work. The solid lines indicate the best fit relation obtained with Eq. \ref{final_ms}. Points and lines are color-coded as a function of time, as indicated in the panel. For clarity, the MS relations are artificially displaced by 0.4 dex from one another. The bottom panel shows the residual distribution as a function of stellar mass. The dashed red line shows the 0 level, corresponding to the best fit value. The red points indicate the mean value of the residuals, while the dotted and dashed-dotted lines indicate the 1 and 3$\sigma$ regions, respectively.}
\label{speagle}
\end{figure}

\section{Fitting the Main Sequence}

In this work, we adopt two approaches to fit the MS relation. The first one follows S14 and consists in looking for the functional form that best describes the MS relation.  As an alternative approach, we investigate the functional form proposed by \cite{2015ApJ...801...80L}, which has the advantage of being expressed with parameters that have a physical description.  In the following procedure, after correcting stellar masses and SFRs with the calibration described above, we take into account the redshift and mass ranges of each study, and we include in the fitting procedure only objects or stacked points actually observed at given mass and redshift without any extrapolation or interpolation. These mass ranges have either been taken directly from the paper in question or estimated based on the data included in the relevant fits, rounded to the nearest $0.1$\,dex after excluding outlying points (see Appendix \ref{selected_ms} for a detailed description of the data taken from each publication).

\subsection{S14 approach}
In this Section, we describe how we fit the MS and retrieve the functional form that best expresses the evolution of the slope and normalization as a function of time. We follow a revised version of the S14 approach. We proceed to fit the evolution of the SFR provided by each individual MS as a function of time:
\begin{equation}
logSFR(t) = \alpha_it + \beta_i,
\label{SFR}
\end{equation}
The choice of fitting as a function of time rather than redshift is mainly practical, as a straightforward linear fit works very well for the time variable, while a more complicated functional form is required as a function of redshift. 

By fitting $\alpha_i$'s and $\beta_i$'s for a grid of M$_{\star}$, as shown in Fig. \ref{linear}, we can derive a function of the form:
\begin{equation}
logSFR(t,logM_{\star}) = \alpha(logM_{\star})t + \beta(logM_{\star}),
\end{equation}
assuming a given parametrization for $\alpha(logM_{\star})$ and $\beta(logM_{\star})$. 

As shown in the right panel of Fig. \ref{linear}, the slope $\alpha(logM_{\star})$ depends linearly on $logM_{\star}$, consistently with S14. Conversely, the best fitting form for $\beta(logM_{\star})$ is not a simple linear dependence but it requires a quadratic form. This is because, differently for nearly all the MS compiled by S14, most of the MS estimates included here find a bending of the MS towards large stellar masses. Thus, the best fitting functions for $\alpha(logM_{\star})$ and $\beta(logM_{\star})$ are, respectively:

\begin{align}
\alpha(logM_{\star}) &= a_0+a_1logM_{\star} \nonumber \\
\beta(logM_{\star}) &= b_0 + b_1logM_{\star}+b_2log^2M_{\star}
\label{ciccio}
\end{align}
which gives:
\begin{align}
logSFR(t,logM_{\star})  &= (a_1t+b_1)logM_{\star} \nonumber \\
								&+b_2log^2M_{\star}+(b_0+a_0t).
\label{final_ms}
\end{align}

Eq. \ref{final_ms} differs from the best fitting function of S14 for the quadratic term $b_2log^2M_{\star}$, which is time-independent. The slope of the MS is driven by the combination of the quadratic term and the linear term that sets the faint end slope, $(a_1t+b_1)logM_{\star}$, which evolves with time. The normalization of the relation depends linearly on time through the term $(b_0+a_0t)$.

\begin{figure*}
\includegraphics[width=\columnwidth]{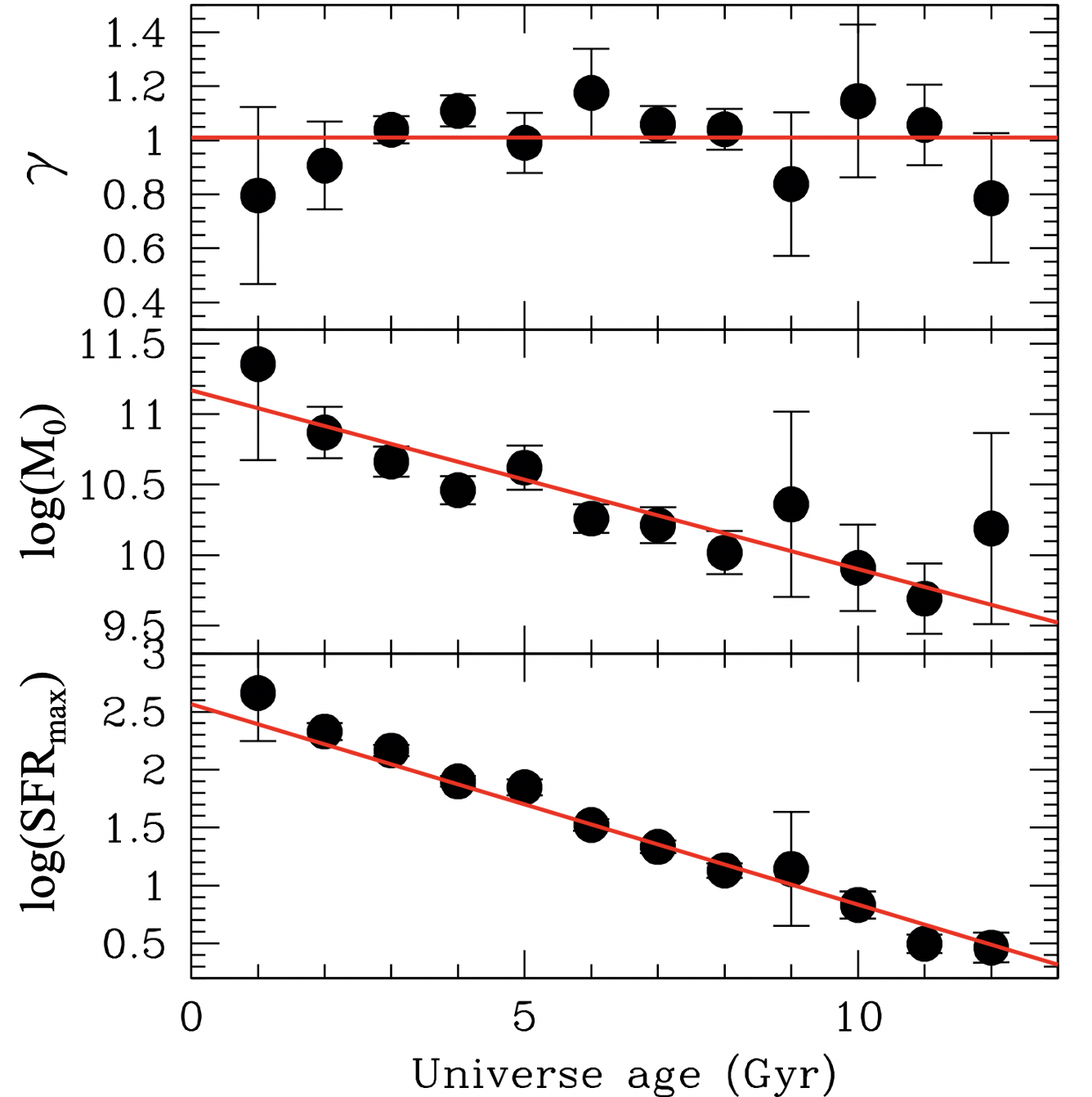}
\includegraphics[width=\columnwidth]{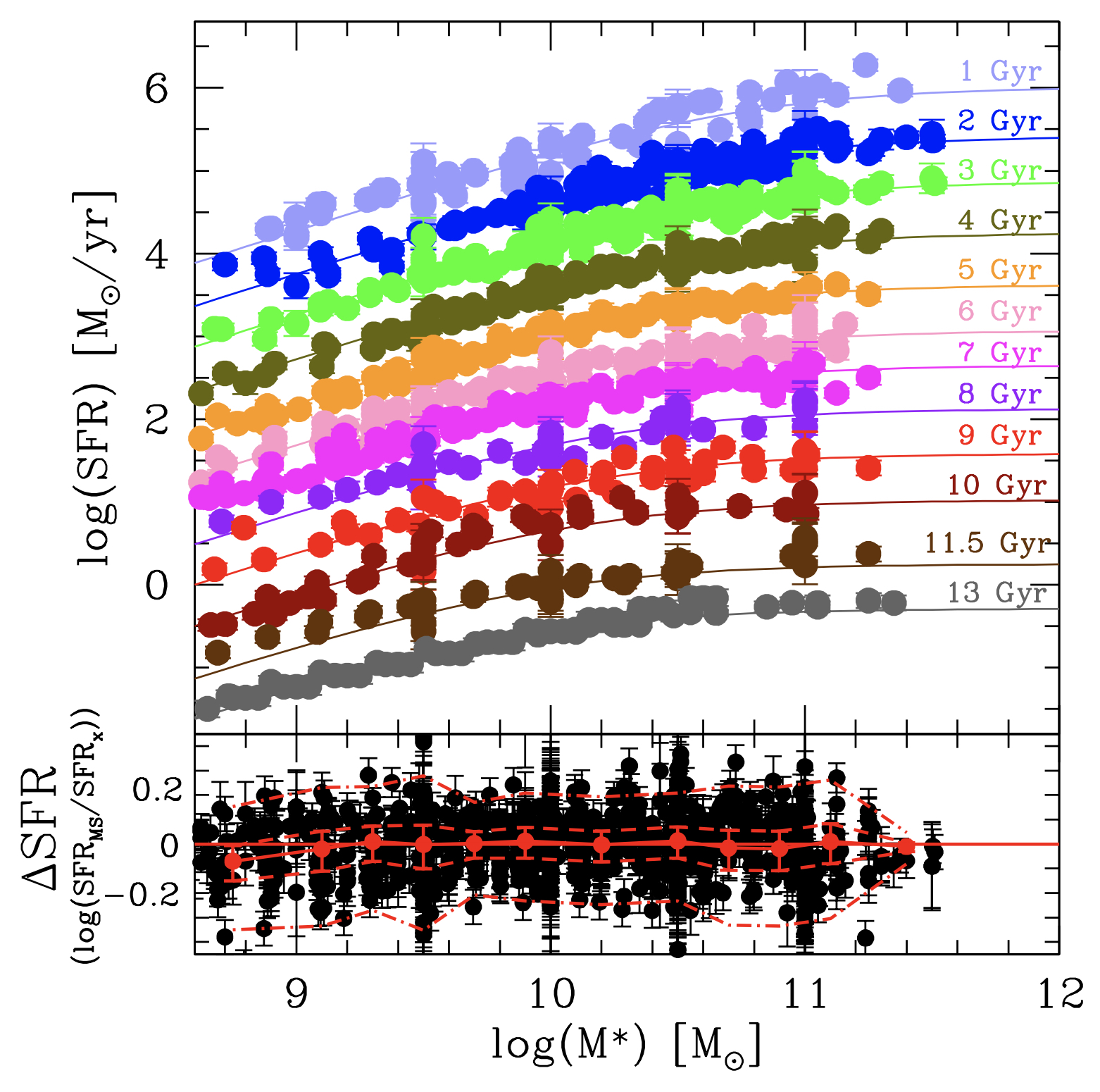}
\caption{{\it{Left panel:}} evolution of $\gamma$ (upper panel), $log(M_0$) (central panel) and $log(SFR_{max})$ (bottom panel) as a function of time. The red solid lines in each panel indicate the best fit relations: a constant equal to 1 for $\gamma$ and the best fit of Eq. \ref{ciccio1} and Eq. \ref{ciccio1a} for $M_0$ and $SFR_{max}$, respectively. {\it{Right panel:}} the upper panel shows the MS shape as a function of the Universe age from 1 to 13 Gyrs. The data points show the MS estimates collected and calibrated in this work. The solid lines indicate the best fit relation obtained with Eq. \ref{Lee_final}. Points and lines are color-coded as a function of time, as indicated in the panel. For clarity, the MS relations are artificially displaced by 0.4 dex from one another. The bottom panel shows the residual distribution as a function of stellar mass. The dashed red line shows the 0 level, corresponding to the best fit value. The red points indicate the mean value of the residuals, while the dotted and dashed-dotted lines indicate the 1 and 3$\sigma$ regions, respectively.}
\label{Lee}
\end{figure*}

We use the functional form retrieved in Eq. \ref{final_ms} to fit the whole MS dataset without binning in stellar masses and time. We limit, however, the fitting procedure to the $10^{8.7}  - 10^{11.3}$ $M_{\odot}$ range, where the MS included in our collection were declared to have high completeness in stellar mass. The best fit parameters of Eq. \ref{ciccio} for $\alpha(logM_{\star})$ and $\beta(logM_{\star})$ are used as first guess for the fitting procedure. The final best-fit parameters are given in Table \ref{fit} and the best-fit MS is shown as a function of time in Fig. \ref{speagle}. The bottom panel shows the residual distribution of the MS estimates with respect to the best fit at a given time and stellar mass. The scatter around the best fit is 0.09 dex. The MS is found to bend towards large stellar masses at all times due to the time-independent quadratic term. The time variation of the faint end slope makes the relation steeper at early epochs. Nevertheless, the evolution of the slope is marginal after the first $\sim$4 Gyrs, consistently with the results of S14, and as found in \cite{2019MNRAS.tmp.2263P}. The normalization of the relation evolves consistently with the results of S14. We point out that it evolves roughly as $(1+z)^3$. However, this is only an approximation, because the SFR(z) can not be expressed accurately as a power law of $(1+z)$. The observed evolution is consistent with previous results in the literature, where there is general consensus on the MS normalization evolving as $(1+z)^{2.8-3}$ at least up to z$\sim4$ \citep[e.g.][]{Sargent2012,2015A&A...575A..74S,2017A&A...605A..70D}. Such evolution has been interpreted as reflecting the evolution of the molecular gas mass density and of the consequent availability of gas supply for the galaxy star formation process \citep{Daddi2008,Daddi2010,Tacconi2010,Tacconi2018}.

\subsection{The MS turn-over}
\label{turnover_m}

The approach of S14 is very effective in providing a good functional form for the MS. However, it is not trivial to physically interpret Eq.\ref{final_ms}. For this reason, we explore a different approach by using the fitting function of \cite{2015ApJ...801...80L}:
\begin{equation}
SFR(M_{\star}) = SFR_{max}/(1+(M_{\star}/M_0)^{-\gamma})
\label{Lee_f}
\end{equation}
Unlike polynomial fits, as Eq. \ref{final_ms}, the parameters of this model allow to quantify the interesting characteristics of the relation between stellar mass and SFR: {\it i)} $\gamma$, the power-law slope at low stellar masses, {\it ii)} ${M}_{0}$, the turnover mass, and {\it iii)} $logSFR_{max}$, the maximum value  of $logSFR$ that the function asymptotically approaches at high stellar masses. 

To capture the evolution of the MS through Eq. \ref{Lee_f}, we first fit the MS in bins of time to estimate the time dependence. This allows to obtain $S_0(t)$, $M_0(t)$ and $\gamma(t)$ to find the best functional form, which includes the time dependence in Eq. \ref{Lee_f}. The left panel of Fig. \ref{Lee} shows the results of the best-fit parameters as a function of time. We find that the exponent $\gamma$ does not show any time dependence and it is consistent with the value $-1$. Both $logSFR_{max}$ and $logM_0$ depend linearly on time. Thus, they can be expressed as:

\begin{equation}
logSFR_{max}(t)=a_0+a_1t
\label{ciccio1}
\end{equation}
and 
\begin{equation}
logM_0(t)=a_2+a_3t
\label{ciccio1a}
\end{equation}

By taking into account such evolution, we can write Eq. \ref{Lee_f} as a function of time as:
\begin{equation}
logSFR(M_{\star},t) = a_0+a_1t-log(1+(M_{\star}/10^{a_2+a_3t})^{-a_4})
\label{Lee_final}
\end{equation}

We use this functional form to fit the MS estimate dataset as a function of mass and time without any binning. The results of the best fitting procedure are indicated in Table \ref{fit}. The right panel of Fig. \ref{Lee} shows the MS data points as a function of time with the best fitting curves. The scatter around the best fit is 0.09 dex, indicating that also Eq. \ref{Lee_final} provides an excellent fitting function for the evolution of the MS. Consistently with the results obtained with the functional form given by Eq. \ref{final_ms}, the MS bends at the high mass end with a turn-over mass that is evolving with time. The exponent $\gamma$ does not evolve with time and it is consistent with the value $1$. This implies that the MS is well represented by the functional form:
\begin{equation}
SFR(M_{\star}) = SFR_{max}(t)/(1+(M_0(t)/M_{\star}))
\label{real_ms}
\end{equation}
which is regulated by only two parameters, $M_0(t)$ and $SFR_{max}(t)$. We estimate the best-fit parameters also for this functional form. The results are included in Table \ref{fit}. The evolution of the MS shape is regulated by the change of the turn-over mass $M_0(t)$. This changes of only 25\% over the past 9-10 Gyrs, and is a factor of 2 larger in the first 3-4 Gyrs. The normalization of the relation, set by $SFR_{max}(t)$ parameter, exhibits the strongest evolution and it is consistent within 1$\sigma$ with the value found for the normalization in Eq. \ref{final_ms}. 

\begin{table*}
\begin{center}
\begin{tabular}{c c c c c c}
    \hline
    \hline \\[-2.5mm]
\multicolumn{2}{c}{Eq. \ref{final_ms}}& \multicolumn{2}{c}{Eq. \ref{Lee_final}} & \multicolumn{2}{c}{Eq. \ref{real_ms}}\\ 
    \hline
    \hline \\[-2.5mm]
 a0 & 0.20$\pm$0.02      & a0 & 2.693$\pm$0.012    & a0 & 2.71$\pm$0.01  \\
 a1 & -0.034$\pm$0.002   & a1 & -0.186$\pm$0.009    & a1 &  -0.186$\pm$0.007\\
 b0 & -26.134$\pm$0.015  & a2 & 10.85$\pm$0.05      & a2 & 10.86$\pm$0.03\\
 b1 & 4.722$\pm$0.012    & a3 & -0.0729$\pm$0.0024    & a3 &  -0.0729$\pm$0.0016  \\
 b2 & -0.1925$\pm$0.0011 & a4 &0.99$\pm$0.01         & a4 & 1\\
  \hline
    \hline \\[-2.5mm]
\end{tabular}
\end{center}
\caption{The table lists the best fit parameters of Eq.\ref{final_ms}, Eq. \ref{Lee_final} in the first 2 columns. The last column lists the best fit parameters of Eq. \ref{real_ms}, which corresponds to Eq. \ref{Lee_final} with the slope $a4=1$ ($\gamma$ in  Eq. 11).}
\label{fit}
\end{table*}

\begin{figure}
\includegraphics[width=\columnwidth]{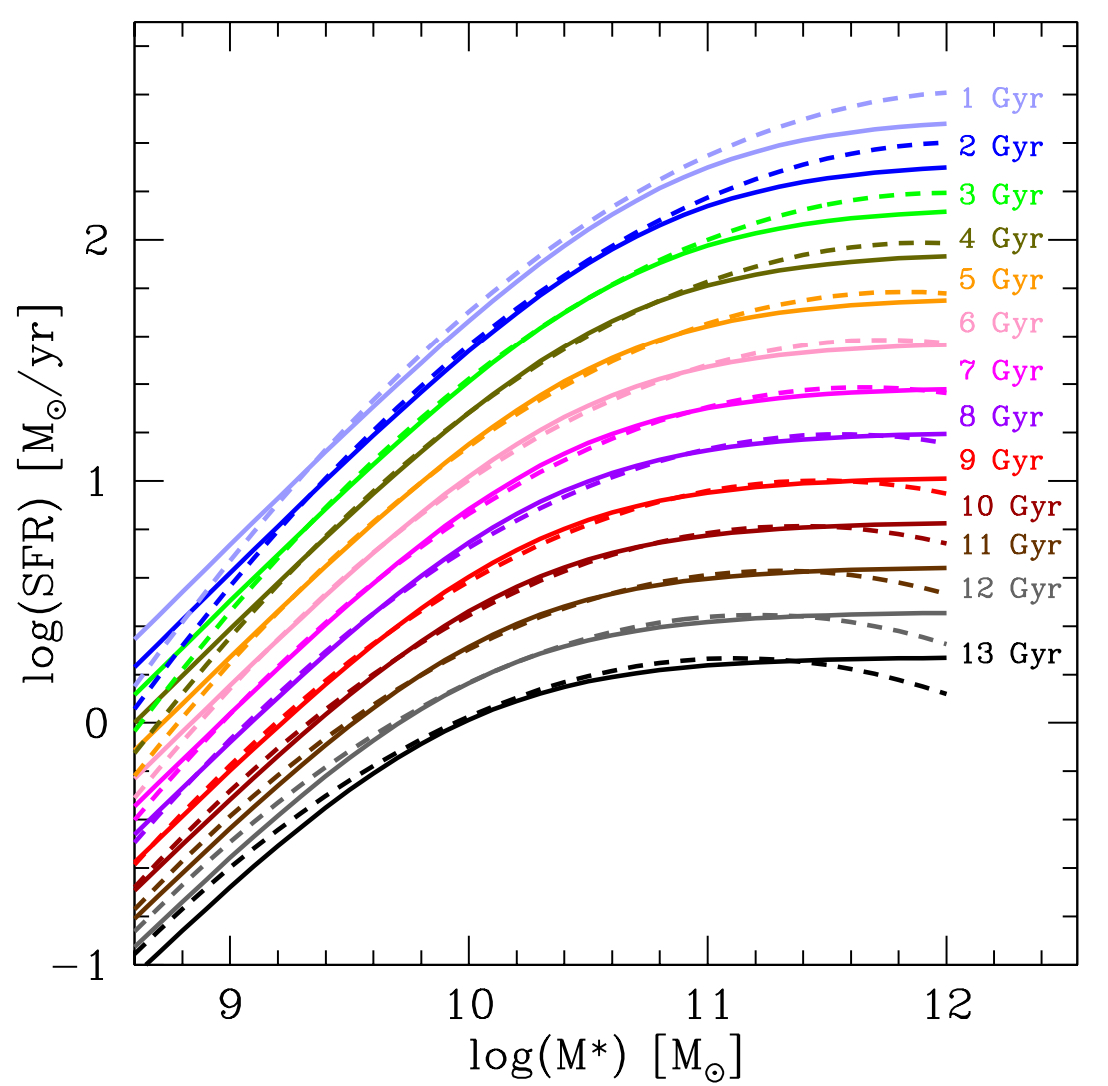}
\caption{Comparison of the two MS best fits as a function of time obtained through Eq. \ref{final_ms} (dotted lines) and Eq. \ref{Lee_final} (solid lines). All lines are color-coded as a function of the Universe age as indicated in the figure  from 1 to 13 Gyrs.}
\label{comparison}
\end{figure}

Fig. \ref{comparison} shows the comparison of the best fit obtained with Eq. \ref{final_ms} and Eq. \ref{Lee_final}. The two fits exhibit a very consistent evolution in normalization as a function of time. We do observe a discrepancy only at the very low mass and high mass ends, in particular at early epochs, where the uncertainty and scatter of the data-points are the highest. Nevertheless, the two fits are consistent with each other within the 1$\sigma$ uncertainty.

\subsection{Towards a physical explanation of the MS shape evolution}

Given Eq. \ref{real_ms}, $M0(t)$ can be interpreted as the mass thresholds between two regimes. At M$_{\star} << M_0(t)$ the sSFR of galaxies is nearly constant as a function of stellar mass and equal to $\sim SFR_{max}(t)/M_0(t)$. At M$_{\star} >> M_0(t)$, the sSFR is progressively suppressed as it is approximated by $SFR_{max}(t)/M_{\star}$. Thus, the turn-over stellar mass separates a regime of constant star formation rate per unit of stellar mass from a regime of SFR suppression. 

\cite{2019MNRAS.483.3213P} point out, exploiting the halo mass catalog of \cite{2007ApJ...671..153Y}, that at z$\sim$0 the region of the MS (within 3$\sigma$ from the relation) is completely dominated by central galaxies. Due to the rather tight correlation between the central galaxy stellar mass and the host halo mass (M$h$), this implies that the SFG mean host halo mass is increasing along the MS with M$_{\star}$. Up to z$\sim$1.3, we are able to check if this holds by exploiting the cosmic web catalog of \cite{2017ApJ...837...16D} in the COSMOS field. This is based on the accurate photometric redshifts of the COSMOS15 catalog \citep{2016ApJS..224...24L} to identify clusters, groups, and filaments and assign a membership probability to each galaxy up to $z\sim1.3$ and down to stellar masses of $10^9$ $M_{\odot}$. For galaxies with a high probability to be in groups and clusters, the catalog provides also a classification in central and satellite galaxies, by identifying as central the most massive system. We use the IR selected galaxy catalog of \cite{2019MNRAS.tmp.2263P} in the COSMOS field, based on the combination of {\it{Herschel}} and {\it{Spitzer}} MIPS data, to check the central galaxy fraction in the MS region in the redshift window explored by \cite{2017ApJ...837...16D}. The SFR of each galaxy is given by the combination of the UV and IR contribution \citep{2019MNRAS.483.3213P}. The MS is identified as the region within 3$\sigma$ from the MS relation given by Eq. \ref{Lee_final} in several redshift bins. We assume $\sigma =0.3$ dex. At all redshifts up to $z=1.3$, the MS region turns out to be dominated by central galaxies, which account for 70\% of the galaxy population. To check whether the MS region is dominated by central galaxies also at z$>$1.3, we use the predictions from the Illustris TNG hydro-dynamical simulation   \citep{2018MNRAS.473.4077P}. In this case, we use as reference the MS determined as in \cite{2019MNRAS.485.4817D} on the same data. Also in this case, as expected, central galaxies account for 70-80\% of the galaxy population at $z > 1.3$. Thus, is it plausible to assume that the MS is dominated by central galaxies also in the distant Universe.

\begin{figure}
\includegraphics[width=\columnwidth]{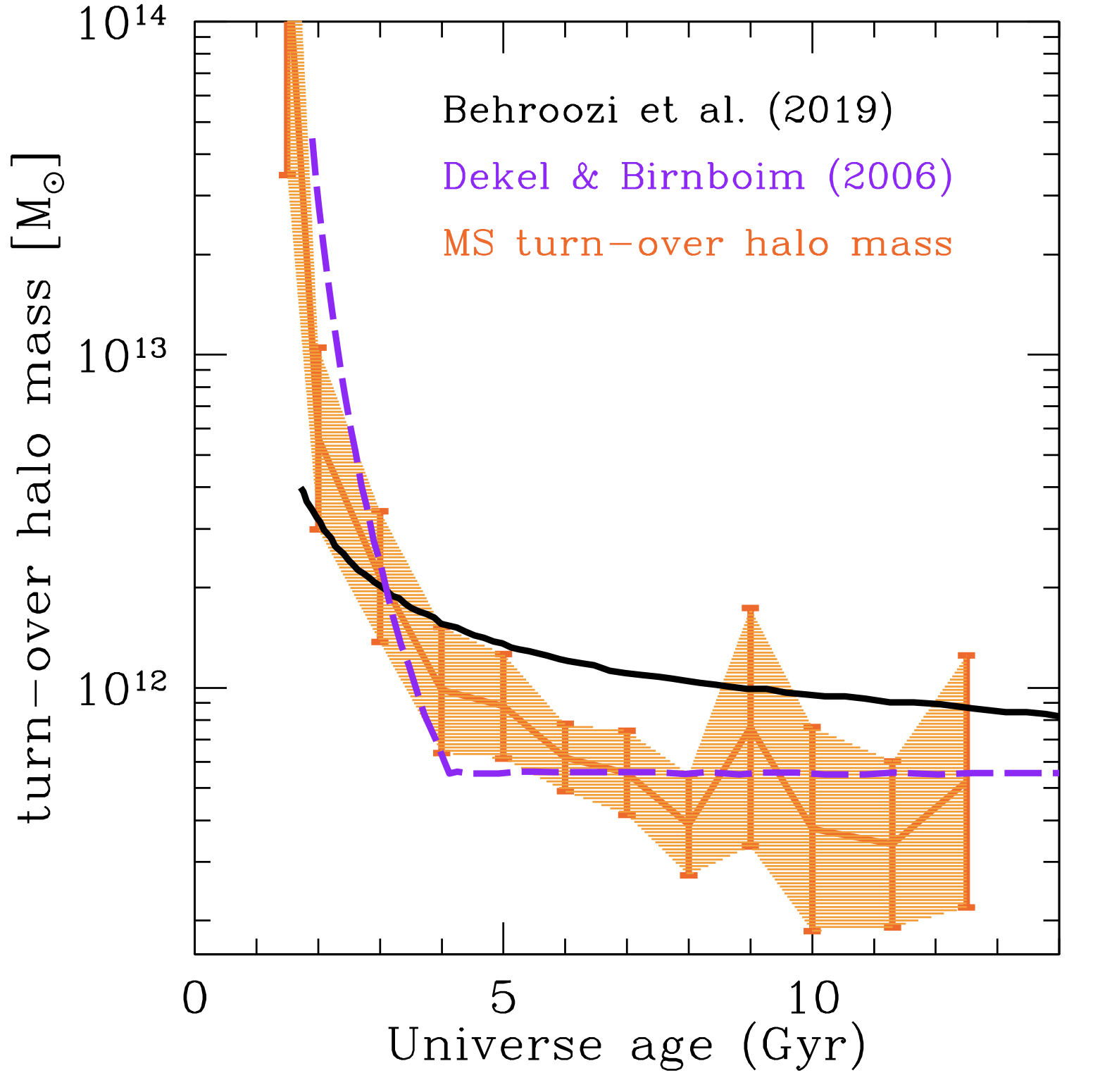}
\caption{Evolution of the host halo turn-over mass (orange line) as a function of the Universe Age. The shaded orange region indicates the 1$\sigma$ uncertainty. This is estimated by combining the error on $M_0(t)$ and the uncertainty in the central galaxy stellar mass and host halo mass correlation of UNIVERSEMACHINE \citep{2019MNRAS.488.3143B}. The solid black line indicates the evolution of the quenching halo mass of \protect\cite{2019MNRAS.488.3143B}. The dashed magenta line shows the evolution of the halo mass threshold between cold and hot accretion regime of \protect\cite{2006MNRAS.368....2D}.}
\label{turnover}
\end{figure}

This aspect is crucial because it allows us to convert the turn-over stellar mass of the SFG MS into a {\it{turn-over host halo mass}}, thanks to the correlation between central galaxy stellar mass and host halo mass \citep{2007ApJ...671..153Y,2013ApJ...770...57B,2019MNRAS.488.3143B}. In particular, we use the results of the empirical model UNIVERSEMACHINE of \cite{2019MNRAS.488.3143B} to convert $M_0(t)$ into host halo turn-over mass $M_{h0}(t)$.
In Fig. \ref{turnover} we plot the evolution of $M_{h0}(t)$ as a function of time together with the halo mass quenching threshold derived in \cite{2019MNRAS.488.3143B}, and the evolution of the transition mass between cold and hot accretion predicted by the theory of mass accretion as in \cite{2006MNRAS.368....2D}.
The halo mass quenching threshold is defined as the halo mass above which the fraction of quenched galaxies is larger than 50\%. The hot/cold transition, instead, is defined as the halo mass at which the cold gas streams coming from the cosmic web filaments are no longer able to penetrate the halo and feed the central galaxy. The curve of \cite{2006MNRAS.368....2D} depends on the halo temperature, but it predicts that at higher redshift ($z> 2.5-3$) cold gas streams are still able to penetrate massive hot halos. Thus, it represents the threshold between hot and cold gas accretion onto the central galaxy. \cite{2019MNRAS.488.3143B} discuss that the disagreement between the empirical and the theoretical predictions might originate from the non-inclusion of the effects of black hole feedback in the treatment of accretion. This, indeed, might play an important role in affecting the thermodynamical conditions of the circum-galactic-medium (CGM) of the central galaxy, mainly by injecting large quantities of energy into it \citep{2017MNRAS.465.3291W,2019MNRAS.tmp.2010N}.

The evolution of $M_{h0}$ with time is remarkably in agreement with the theoretical model of \cite{2006MNRAS.368....2D}. The evolution of $M_{h0}(t)$ decreases steeply in the first 4-5 Gyrs of the Universe and it reaches a plateau afterward. This would suggest that the turn-over mass of the MS might be indicative of the transition between an environment that efficiently sustains the star formation process of the central galaxies, e.g. through cold gas streams, to one which is hostile to the star formation process due to the suppression of the feeding mechanism of the central galaxy \citep[see also][]{Tacchella2016,Daddi2022}. This suppression is likely maintained in the hot environments, as suggested by nearly all the most sophisticated hydrodynamical simulations, by the interplay between the hot gas in massive halos and central black hole feedback \citep{2015ApJ...808L..30V,2017ApJ...845...80V,2017MNRAS.465.3291W,2019MNRAS.tmp.2010N}.

\section{Comparison with previous results}

\begin{figure}
\includegraphics[width=\columnwidth]{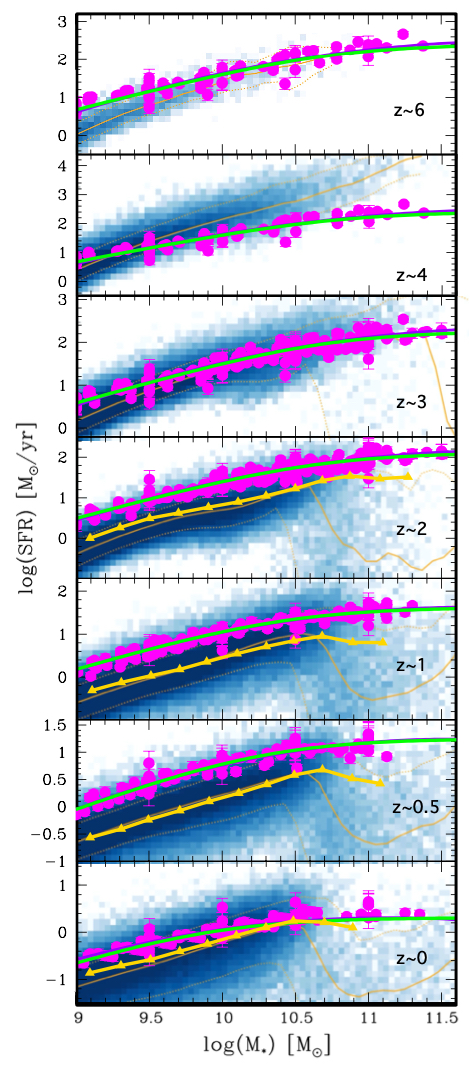}
\caption{MS of star-forming galaxies in several redshift bins from $z\sim6$ to $z\sim0$. The shaded blue region in each panel indicates the distribution of simulated Illustris TNG galaxies in the SFR-stellar mass plane. The blue scale is according to the number density of galaxies. The orange line shows the running median in bins of stellar mass for the simulated galaxies, while the connected yellow triangles show the relation of \protect\cite{2019MNRAS.485.4817D} limited to the UVJ selected galaxies in Illustris TNG. The magenta points show the data collected in this paper. The green and purple solid lines indicate our best fit according to Eq. \ref{final_ms} and \ref{Lee_final}.}
\label{illustris}
\end{figure}

In this Section, we compare our results on the evolution of the MS with previous observational and theoretical results.

We first compare our results to the ones of S14, keeping in mind that the analysis of S14 is limited to the stellar mass range $10^{9.7}-10^{11}$ $M_{\odot}$. In Fig. \ref{illustris} we plot in magenta the homogenized collection of MS relations in seven different redshift bins. The green solid line marks our best-fit estimates at the given redshift, as expressed in Eq. \ref{final_ms}, while the green line indicates the best fit of Eq. \ref{Lee_final}. The MS relation of S14 is perfectly overlapping our relation in their limited stellar mass range. S14 claim that they do not find a bending of the MS at the high mass end. However, we point out that this could be due to the limited stellar mass range considered in the fitting procedure. Indeed, while a single power law might be a good approximation over this range, the extrapolation to larger masses would largely disagree with the data above $10^{11}$ $M_{\odot}$.

The bending of the MS has been largely discussed in the literature and the redshift and stellar mass at which the relation bends vary largely among the different publications included in our collection, from an MS bending only at $z < 2$ \citep[e.g.][]{2018ApJ...853..131L}, to a correlation that becomes a power law at $z > 2$ \citep{2015A&A...575A..74S,2016ApJ...817..118T}, or an MS constant in shape and evolving only in normalization \citep{2012ApJ...754L..29W,2014ApJ...795..104W,2015ApJ...801...80L}. In this work, we find that the MS bends at all redshifts. However, the bending happens above very large stellar masses at early epochs ($z > 3$). Works limited to lower stellar masses or to relatively small deep fields can not capture this feature. In addition, some of the previous works focus on different aspects, from the analysis of the low mass slope of the relation in a few cases \citep[e.g.][]{2012ApJ...754L..29W,2014ApJ...795..104W}, to the shape and normalization at the high mass end in others \citep[e.g.][]{sherman2021}. We point out, though, that the bending observed in the MS estimated in this work is the result of the combination of all these different estimates, which all tend to be in agreement within a relatively small scatter (0.08 dex), when brought to a common framework. This, perhaps, might suggest that the reported discrepancies are mainly due to possible biases introduced by a limited stellar mass range, the limited volume of the studied deep fields, possible low number statistics, systematics due to SFR indicator, or a different fitting function. Thus, the approach suggested by S14 and implemented here has the potential of overcoming the limitations of the individual analysis and offers a broader and more complete view of the evolution of the MS over a much larger time interval and stellar mass range.

\begin{figure}
\includegraphics[width=\columnwidth]{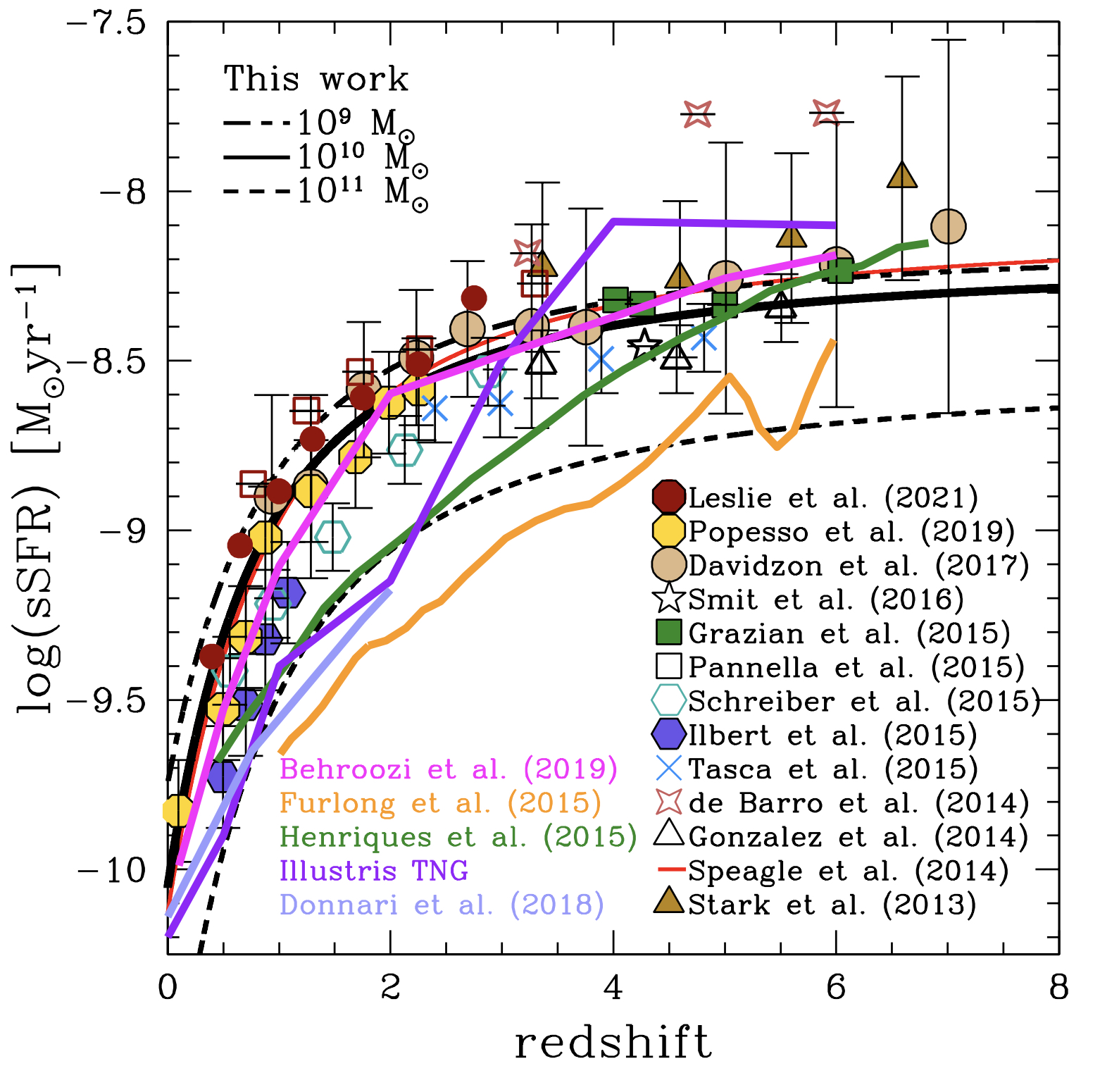}
\caption{Evolution of sSFR as a function of redshift. The black lines show the sSFR evolution given by the best fit of Eq. \ref{Lee_final} at three stellar mass values: $10^9$ (dot-dashed line), $10^{10}$ (solid line) and $10^{11}$ $M_{\odot}$ (dashed line). We provide the same relation for the S14 best fit at $10^{10}$ (red line). We plot for comparison the data of several works in literature with symbols indicated in the figure. The yellow, green, purple and light purple solid lines show the predictions of EAGLE simulation of \protect\cite{2015MNRAS.450.4486F}, The MUNICH simulation of \protect\cite{2015MNRAS.451.2663H}, the Illustris TNG as estimated here and the one of \protect\cite{2019MNRAS.485.4817D}, respectively. The magenta solid line shows the best fitting model of \protect\cite{2019MNRAS.488.3143B}.}
\label{ssfr}
\end{figure}

Fig. \ref{illustris} shows also the comparison between our results and the predictions of the Illustris TNG300 simulation \citep{2018MNRAS.473.4077P}. TNG300 is the largest volume simulated in the suite of Illustris TNG hydrodynamical simulations. The choice of TNG300 is driven by the need to sample a sufficiently large volume to capture the rare giant star-forming galaxies at the high mass end of the MS. The SFRs are averaged over 200 Myr and measured within a physical aperture of $2 R_{star}$, where $R_{star}$ is the stellar half mass radius (see also \citealt{2019MNRAS.485.4817D}). The shaded region in Fig. \ref{illustris} shows the distribution of the IllustrisTNG galaxies, color-coded according to the galaxy number density in bins of SFR and $M_{\star}$. The IllustrisTNG MS is estimated as a running mean as a function of $M_{\star}$ and it is indicated by the orange-yellow line. We also show in yellow the MS of \cite{2019MNRAS.485.4817D}, which is estimated up to $z\sim2$ by mocking the UVJ selection of \cite{2012ApJ...754L..29W}. The two estimates agree remarkably well up to stellar masses of $\sim 10^{10.5}$ $M_{\odot}$ and diverge at larger stellar masses, where the UVJ selection of \cite{2019MNRAS.485.4817D} exclude most of the low SFR systems.

It is interesting to notice that a bending of the predicted MS is observed also in Illustris TNG at least up to z$\sim 3$. This was already pointed out by \citet{2019MNRAS.485.4817D} at $z < 2$. At higher redshift, the relation is steeper than the observed MS, but this could be due to a lack of massive star-forming galaxies in the simulations, as pointed out below.

We notice a good agreement between the IllustrisTNG predictions and our results at early epochs ($z\sim6$). The observed MS lies over the simulated relation. However, no giant star-forming galaxies are observed at all in the TNG300 volume at $z\sim6$ above $10^{10.5}$ $M_{\odot}$. The observations, instead, indicate that such galaxies populate the high mass end of the relation, in agreement with the results of the GSMF of active galaxies of \cite{2017A&A...605A..70D}. This is also confirmed by the recent JWST discoveries of very massive blue galaxies in the distant Universe \citep[e.g.][]{Santini2022,Finkelstein2022,Castellano2022,Donnan2022}

Massive star-forming galaxies appear in the TNG simulation only between $z\sim4$ to $\sim3$, with a larger SFR compared to observations, in particular at $z\sim 4$. At this epoch the predicted and observed low mass slope (below stellar masses of $10^{10.5}$ $M_{\odot}$) are in agreement within 1$\sigma$. At later epochs ($z < 2$), the predicted MS is systematically below the observations by $\sim 0.5$ dex, according to the results of \cite{2019MNRAS.485.4817D}. Observations and predictions are again in agreement at $z\sim 0$ up to $\sim 5\times 10^{10}$ $M_{\odot}$. Above this mass, no more SFGs are detected in the simulations at odds with observations.

The tension between observations and simulations in the evolution of the MS is not new and has been extensively discussed in the literature \citep[see, for instance,][]{Katsianis2020,Nelson2021}. From the observational point of view, the use of sophisticated SED fitting codes, such as the {\it{Prospector}} or its incarnation of {\it{Prospector-$\alpha$}} used in \citet{Leja2022}, might apparently solve this issue. Indeed, adding an extra population of old stars might justify an overestimation of the previous {\it{Spitzer}} and {\it{Herschel}} based SFR estimates and reconcile ad hoc the observed and simulated MS in specific redshift bins. However, it is worth pointing out that such an approach might simply move the tension somewhere else. Indeed, the stellar masses provided by {\it{Prospector}} increase considerably with respect to previous works. This increase would shift the stellar mass function leading to a substantial disagreement with the predicted ones. These, indeed, are overall in agreement with previous measurements, as shown, for instance, in \citet{pille2018} for Illustris TNG and in the detailed comparison of \citet{Thorne2021}. 

It is, perhaps, more interesting to point out that the largest disagreement between the observed and the simulated MS is in the distribution of the star-forming galaxy population along the MS. Fig. \ref{illustris} shows quite clearly that, in the observations, star-forming galaxies with stellar masses above $\sim 5\times 10^{10}$ $M_{\odot}$ exhibit a much slower evolution than predicted by the simulations. Indeed, they should have formed by $z\sim 4$, while in Illustris TNG they do not appear at all. Furthermore, the high mass end of the MS, above this mass threshold, is almost completely evacuated by $z\sim 0$ in the simulation. Instead, many works in the literature, based on the SDSS galaxy spectroscopic sample or the WISE survey, show clearly that this region of the SFR-stellar mass plane is highly populated up to stellar masses of $3\times 10^{11}$ $M_{\odot} $\citep[see, for instance,][for a collection of the available datasets in the local Universe]{2019MNRAS.483.3213P}. Such discrepancy can not be ascribed to an overestimation of the observed SFR, as recently proposed \citep{Leja2022}. Indeed, this would simply lead to a discrepancy of the MS normalization rather than to the lack of the massive SFG sub-population. Thus, it is, perhaps, more likely that the predicted fast evolution of massive SFG is due to an over-efficient SF quenching process in simulations. We point out that the use of the larger stellar masses estimated by {\it{Prospector}} would increase the observed discrepancy at the high mass end, not only in the local Universe but also at higher redshift ($0.5 < z < 2$).

We provide in Fig. \ref{ssfr} a comparison of the sSFR evolution with simulations, as well as other works in literature. We plot the evolution of the sSFR derived from the best fit of Eq. \ref{Lee_final} computed in three different stellar mass bins: $10^{9}$ (black dot-dashed line), $10^{10}$ (black solid line) and $10^{11}$ $M_{\odot}$ (black dashed line). The red line shows the result of S14  at $10^{10}$ $M_{\odot}$, which is in agreement with our result at the same stellar mass. The prediction of IllustrisTNG, up to $z\sim 6$, those from \cite{2019MNRAS.485.4817D}, up to $z\sim 2$, as well as the best fit of the empirical model UNIVERSEMACHINE \citep{2019MNRAS.488.3143B}, all computed for M$_{\star} =10^{10}$ $M_{\odot}$, are shown in purple, lavender and magenta, respectively. The simulated data of Illustris TNG are in agreement with observations at high redshift and in the local Universe, but the slope of the predicted sSFR evolution is steeper than observations, as pointed out in the previous paragraph. The redshift evolution of sSFR obtained from \cite{2019MNRAS.488.3143B} is perfectly overlapping with observations at the same stellar mass. Finally, we show in Fig. \ref{ssfr} the redshift evolution of the sSFR obtained from the EAGLE hydrodynamical simulation \citep{2015MNRAS.450.4486F} and the Munich simulation of \citet{2015MNRAS.451.2663H}. Both predictions lie below the observations and exhibit a steeper relation. However,  we point out that, as discussed by \cite{2018ApJ...852..107D}, such comparison is complicated by the fact that simulated galaxies are not selected to be SFGs. Thus, the slope and normalization of the sSFR-redshift relation is biased by quiescent galaxies, which might have a faster evolution.

{We conclude that the star formation activity of simulated galaxies is declining much faster than in observations between redshift $\sim3 $ and $\sim 0$. This leads to an MS normalization too low with respect to observations in the same redshift window. The disagreement is more significant for the most massive systems above $10^{11}$ $M_{\odot}$, which completely disappear from the simulated MS in the local Universe, at odds with observations. We do not find evidence for a clear overestimation or underestimation of the SFRs in specific redshift bins, that might solve the tension.  As discussed in \cite{2019MNRAS.485.4817D}, the observed discrepancy might point to some fundamental limitations in our understanding of the processes governing the star formation evolution in galaxies, such as the role of cold and hot accretion and of supernovae and black hole feedback.

\section{Summary and Conclusion}
We compile a collection of the most important publications regarding the evolution of the MS of star forming galaxies in the widest range of redshift ($0 < z < 6$), stellar mass ($10^{8.5}-10^{11.5}$ $M_{\odot}$) and SFR ($0-500$ $M_{\odot}yr^{-1}$) ever probed. We convert all observations to a common calibration to check for consistency in the literature estimates and to study the evolution of the relation with different approaches. We find a remarkably good agreement between the different estimates at any stellar mass and time. The resulting MS exhibits a curvature towards the high stellar masses, which is slowly evolving with time. We provide two functional forms, which take into account the time evolution of the MS normalization and slope. Following the approach of S14, we estimate the best polynomial fitting form in the logSFR$-$logM$_{\star}$ space, by studying the evolution with time of the logSFR at fixed stellar mass. A second-order polynomial form is well representing the relation, with a non-evolving quadratic term. The normalization is evolving as a power law of the Universe age. The slow evolution of the linear term reproduces the steepening of the relation towards the first 3-4 Gyrs of the Universe. We provide, as alternative fitting form, the one of \cite{2015ApJ...801...80L}, which has the advantage of being expressed by physical parameters. These are the low mass slope, the normalization, and the turn-over mass ($M_0(t)$). While the slope does not evolve with time, normalization and turn-over mass evolve as a power law of the Universe age. The turn-over mass, in particular, determines the MS shape. It marginally evolves with time, and it is responsible for the steepening of the relation towards $z\sim4-6$. At stellar masses below $M_0(t)$, SFGs have a constant sSFR, while above $M_0(t)$ the sSFR is suppressed. As the MS region is dominated by central galaxies, we use the relation between central galaxy stellar mass and host halo mass to convert $M_0(t)$ into a "turn-over host halo mass". We find that its evolution is remarkably consistent with the one of the halo mass threshold between cold and hot accretion regimes predicted by \citet{2006MNRAS.368....2D}. This might indicate that $M_0(t)$ defines the transition between an environment able to sustain the SF process of the central galaxy, for instance through cold gas streams as predicted by \cite{2006MNRAS.368....2D}, to a regime hostile to the same process. The latter might be the result of the interplay between the hot gas in massive halos and the black hole feedback generated by the central galaxy itself. 

The comparison of our results with the state-of-the-art hydrodynamical simulations shows that the simulated MS shape is qualitatively consistent with the observations. However, the normalization of the simulated relation is systematically lower by at least 0.2 to 0.5 dex with respect to observations at $0.5 < z < 3$. As a consequence, the sSFR of SFGs evolves more rapidly in simulated galaxies with respect to the observed ones. We do not find clear evidence for an overestimation of the observed SFR at $0.5 < z < 3$ with respect to simulated SFR. This might suggest that the feedback implemented in the simulations is likely too efficient in suppressing the star formation activity of galactic systems, in particular, above $5\times 10^{10}$ $M_{\odot}$ \citep[see also][]{Katsianis2021,Corcho-Caballero2021}.

\section*{Acknowledgements}
We would like to deeply thank the referee, A. Katsianis, for helping in taking a different perspective and for the enormous help in catching up with the literature. This paper was ready for submission at the end of 2019, but P.P. had to stop working for about 2 years because of the pandemic. The referee helped significantly in  improving the manuscript and we consider this as one of the most positive and constructive experiences of a review process. 

\section*{Data availability}
The data underlying this article are publicly available in the articles listed in Table 1 and in their online supplementary material.



\bibliographystyle{mnras}
\bibliography{laura1} 





\appendix
\section{The selected MS}
\label{selected_ms}
Here we list and describe the MS estimates included in the analysis and the data they are based on:
\begin{itemize}
    \item \cite{2014ApJS..214...15S} do not provide the collection of data used for calibrating the average MS estimates but only the best fits. A large number of best fits are provided based on different sub-samples, depending on the star-forming galaxy selection method, and on the range of stellar masses and cosmic time included in the fit. We use in this paper the fit n. 64 based on a "mixed selection", which S14 consider as more inclusive of the SFG population (see Section \ref{selection_effect} for a detailed discussion). The fit is restricted to stellar masses between $10^{9.7}$ $M_{\odot}$ and $10^{11}$ $M_{\odot}$ and it excludes the first 2 Gyrs of the Universe Age. In order to properly populate the S14 MS given by the fit n. 64, we use the data of Fig n. 4, which shows the value of the SFR as a function of time in 4 stellar mass bins. We distribute randomly the data within the stellar mass bins and use the time information to estimate the SFR as a function of stellar mass and time in the range $10^{9.7}-10^{11}$ $M_{\odot}$ and 2-12 Gyrs of Universe age. The SFRs obtained in this way are calibrated to the KE12 calibration and a Kroupa IMF.
    
    \item \cite{2014MNRAS.443...19R} provide the MS estimate of BzK selected galaxies in the redshift bin $1.4 < z < 2.5$ in the COSMOS field, on the basis of various SFR indicators including UV emission, $H\alpha$ emission, MIR and FIR emission. In this analysis, we use the four data points obtained through the stacking analysis of the BzK sample in the COSMOS PACS maps in the stellar mass range $10^{10}-10^{11.5}$ $M_{\odot}$. The SFR and stellar masses are obtained with a Salpeter IMF and are corrected to the KE12 calibration and a Kroupa IMF.
    
    \item \cite{2014MNRAS.437.1268H} provide the MS based on the stacking UV selected SFGs in bins of FUV luminosity in the HerMES maps \citep{2012MNRAS.424.1614O}. The derived SFR is based on the combination of NUV and IR luminosities, with a Chabrier IMF. The MS is estimated between $10^{9.5}$ and $10^{11.3}$ $M_{\odot}$ at $z\sim 1.5$, $z\sim 3$ and $z\sim 4$. We include in our analysis the stacked points at the observed stellar masses and redshift given in the paper after correcting them to the KE12 calibration. 
    
    \item \cite{2014ApJ...795..104W} provide the MS based on stacking of UVJ selected star-forming galaxies on {\it{Spitzer}} MIPS 24 $\mu$m data in the 3D-HST CANDELS fields. The derived SFR is based on the combination of NUV and IR luminosities, with a Chabrier IMF. The considered redshift range is 0.5-2.5 and the MS is estimated in the following stellar mass and redshift ranges: $10^{8.5}-10^{11.2}$ $M_{\odot}$ at $0.5< z < 1$, $10^{9}-10^{11.3}$ $M_{\odot}$ at $1 < z < 1.5$, $10^{9.2}-10^{11.5}$ $M_{\odot}$ at $1.5 < z <2$ and $10^{9.3}-10^{11.5}$ $M_{\odot}$ at $2 < z <2.5$. We include in our analysis the stacked points at the observed stellar masses and redshift given in the paper, recalbrated to KE12.
    
    \item \cite{2015ApJS..219....8C} provide SFR based on SED fitting results of MAGHPHYS code, from GALEX to WISE data in the local Universe. As already reported by \cite{2019MNRAS.483.3213P}, the catalog provides SFRs underestimated with respect to the H$\alpha$ based and IR-based SFR estimates. \cite{2016ApJS..227....2S} discuss that this is likely due to the fact that the MAGHPHYS SED fitting results are mostly driven by the higher SNR 12 $\mu$m WISE data-point than the low SNR 22 $\mu$m WISE data, leading to artificially low SFR. \cite{2015ApJS..219....8C} report that their MS at $z < 0.1$ is systematically lower (0.15 dex) with respect to other H$\alpha$ and IR based MS. We estimate the MS by using the best fit relation in stellar mass bins of 0.15 dex to take into account stellar mass uncertainties and after correcting the SFRs for the 0.15 dex offset, which brings them in agreement to the Ke12 calibration. 
    
    \item \cite{2015ApJ...801...80L} use a ladder of SFR indicators to study the MS in the COSMOS field at $0.3< z <1.3$. Due to the flux limit of the {\it{Spitzer}} MIPS 24 $\mu$m and {\it{Herschel}} PACS and SPIRE catalogs, the combination of NUV and IR luminosities provides the SFR only for highly star-forming objects (starburst and massive SFGs). For less active or less dusty objects they use dust-corrected NUV luminosities (see Fig. 3 of the cited paper for the contribution of any SFR indicator as a function of stellar mass). Star-forming galaxies are selected according to their (M$_{NUV}$-M$_R$) $-$ (M$_R$-M$_J$) colors with a Chabrier IMF. The MS is estimated as the median SFR value in equally populated bins of stellar mass in the following ranges: $10^{8.5}-10^{11.}$ $M_{\odot}$ at $z=0.36$, $10^{9}-10^{11.}$ $M_{\odot}$ at $z=0.55$, $10^{9}-10^{11.}$ $M_{\odot}$ at $z=0.70$, and between $10^{9.3}$  and $10^{11.2}$ $M_{\odot}$ at $z=0.85$, $z=0.99$ and $z=1.19$. We include in our analysis the median points at the observed stellar masses and redshift, after correcting the median into the mean value and to the KE12 calibration.
    
    \item \cite{2015A&A...579A...2I} use the flux-limited sample of {\it{Spitzer}} MIPS 24 $\mu$m and {\it{Herschel}} PACS and SPIRE of the COSMOS field to study the distribution of galaxies in the SFR-stellar mass plane. Galaxies are further selected according to their (M$_{NUV}$-M$_R$) $-$ (M$_R$-M$_J$) colors. The SFR is estimated with a combination of NUV and IR luminosities. The MS is identified as the peak of the SFR distribution at several stellar mass bins at stellar masses above $10^{10}$ $M_{\odot}$ and in the redshift range $0.2 < z < 1.4$. This corresponds to the median SFR in a log-normal distribution. We include in our analysis the median points at the observed stellar masses and redshift given in the paper, after correcting the median to the mean and to the KE12 calibration.
    
    \item \cite{2015A&A...581A..54T} use the spectroscopically selected sample of the VIMOS Ultra Deep Survey (VUDS, \citealt{2015A&A...576A..79L}). Galaxies are selected from a combination of photometric redshifts, as well as from color selection  criteria  like  LBG,  combined with a  flux limit $22.5 < i_{AB} <25$. The SFR is derived via SED fitting techniques with {\it{Le Phare}}, assuming the BC03 SPS model, a Chabrier IMF and a \cite{2000ApJ...533..682C} extinction law. The MS is provided in the following stellar mass and redshift ranges: $10^{7.5}-10^{10}$ $M_{\odot}$ at $0 < z < 0.7$, $10^{8.5}-10^{11}$ $M_{\odot}$ at $0.7 < z < 1.5$, $10^{9}-10^{11}$ $M_{\odot}$ at $1.5 < z < 2.5$ and at $2.5 < z < 3.5$, $10^{9.3}-10^{10.6}$ $M_{\odot}$ at $3.5 < z < 4.5$ and $10^{9}-10^{11}$ $M_{\odot}$ at $z > 4.5$. We include in our analysis the mean points at the observed stellar masses and redshift, corrected to the KE12 calibration.

    \item \citet{Salmon2015} study the evolution of the slope and scatter of the SFR–stellar mass relation for galaxies at $3.5 < z < 6.5$ using multi-wavelength photometry in GOODS-N from the Cosmic Assembly Near-infrared Deep Extragalactic Legacy Survey (CANDELS) and Spitzer Extended Deep Survey. They use an updated, Bayesian spectral-energy distribution fitting method that incorporates effects of nebular line emission, star formation histories that are constant or rising with time, and different dust attenuation prescriptions (starburst and Small Magellanic Cloud). They use a modified version of \citet{2003MNRAS.344.1000B} stellar population synthesis models, and a Salpeter IMF.
    
    \item \cite{2015ApJ...801L..29R} provide a fit of the local MS at $z < 0.085$ in the stellar mass range $10^{9}-10^{10.5}$ $M_{\odot}$. The SFR is based on dust-corrected SDSS H$\alpha$ fluxed for systems classified as star forming in the BPT diagram and through the D4000 break for non-active systems or AGN hosts \citep{2004MNRAS.351.1151B}. SFR and stellar masses are estimated with a Chabrier IMF. The MS is identified as the peak of the SFR distribution at fixed stellar mass in the SFR-stellar mass plane. This corresponds to the median SFR in a log-normal distribution. We estimate the MS by using the best fit relation in stellar mass bins of 0.15 dex to take into account stellar mass uncertainties. The MS estimates are included after correcting the median into the mean value and to the KE12 calibration. 
    
    \item \cite{2015A&A...575A..74S} perform a stacking analysis of UVJ selected galaxies in the deep {\it{Herschel}} PACS maps of the CANDELS fields. The mean IR luminosity derived for the stacks is combined with the mean FUV luminosity to derive the SFR. SFR and stellar masses are estimated with a Salpeter IMF. The MS is given in the following stellar mass and redshift ranges: $10^{9}-10^{11}$ $M_{\odot}$ at $z\sim 0.5$, $z\sim 1$ and $z\sim1.6$, $10^{9.8}-10^{11}$ $M_{\odot}$ at $z\sim 2$ and $z\sim 3$. We include in our analysis the stacked points at the observed stellar masses and redshift, corrected to the KE12 calibration.

    \item \cite{delosreyes2015} use $\sim$300 H$\alpha$-selected  galaxies at $z \sim 0.8$ to study the MS. They use deep optical spectra obtained with the IMACS spectrograph at the Magellan telescope to measure strong oxygen lines. They combine spectral information with rest-frame UV-to-optical imaging, which allows them to determine stellar masses and dust attenuation corrections, and H$\alpha$ narrow-band imaging, which provides a robust measurement of the instantaneous SFR. The SEDs are fit with a library of \citet{2003MNRAS.344.1000B} stellar population synthesis models. The model libraries are built with a wide range of SFHs and metallicities, as described in \citet{2007ApJS..173..267S}, and updated \citet{daCunha2008}. Each model is attenuated according to the prescription of \citet{2000ApJ...539..718C}. The dust attenuation in the SED fitting is mainly constrained by the UV slope. 
    
    Their sample spans stellar masses of $\sim 10^{9}–6 \times 10^{11}$ $M_{\odot}$. They find consistency with previous results about the MS at similar redshift. We do not implement any correction in stellar mass and SFR.
    
    \item \cite{2016MNRAS.455.2839E} provide the MS at $z < 1.1$. The SFR is based on the flux-limited sample of {\it{Spitzer}} MIPS 24 $\mu$m and {\it{Herschel}} PACS and SPIRE available in the ECDFS and COSMOS fields. The SFR is derived from the far-IR flux with a Chabrier IMF. The MS is retrieved via $\sigma$-clipping. It is estimated in the stellar mass range $10^9-10^{11.5}$ at $0.15 <z < 0.5$ and $0.5 < z < 1.1$. We include in our analysis the stacked points at the observed stellar masses and redshift, corrected to the KE12 calibration.
    
    \item \cite{2016ApJ...817..118T} perform a stacking analysis of UVJ selected galaxies in the deep {\it{Herschel}} maps of the CANDELS fields. The mean IR luminosity derived for the stacks is combined with the mean NUV luminosity to derive the SFR. SFR and stellar masses are estimated with a Chabrier IMF. The MS is given in the following stellar mass and redshift ranges: $10^{8.5}-10^{11.2}$ $M_{\odot}$ at $z\sim 0.6$, $z\sim 0.9$, $z\sim1.1$ and $z\sim1.4$, $10^{8.7}-10^{11.2}$ $M_{\odot}$ at $z\sim 1.7$, $10^{9}-10^{11.2}$ $M_{\odot}$ at $z\sim 2.25$ and $z\sim 2.75$, $10^{9.5}-10^{11.5}$ $M_{\odot}$ at $z\sim 3.5$.  We include in our analysis the stacked points at the observed stellar masses and redshift, corrected to the KE12 calibration.
    
    \item \cite{2017ApJ...847...76S} derive the MS in the ultra-deep Hubble Space Telescope Frontier fields. The MS is derived via SED fitting with a Salpeter IMF, BC03 SPS model and a \cite{2000ApJ...533..682C} extinction law. The MS location is derived via $\sigma$-clipping at the following stellar masses and redshift ranges: $10^8-10^{10.6}$ $M_{\odot}$ at $1.3 <z < 2$,  $10^8-10^{11}$ $M_{\odot}$ at $2 <z < 3$ and $3<z<4$, $10^{8}-10^{11}$ at $4<z <5$, and $10^{8.6}-10^{11}$ $M_{\odot}$ at $5 < z < 6$. We estimate the MS by using the best fit relations in stellar mass bins of 0.25 dex to take into account stellar mass uncertainties. We correct the MS to the Kroupa IMF.
    
    \item \cite{2016ApJ...820L...1K} utilize  photometry  in  the Hubble Ultradeep  Field  (HUDF12)  and  Ultraviolet  Ultra  Deep  Field (UVUDF)  campaigns  and  CANDELS/GOODS-S to estimate the SFR via SED fitting. They use a Salpeter IMF, BC03 SPS model and a \cite{2000ApJ...533..682C} extinction law. The MS is identified via $\sigma$-clipping between $10^7$ and $10^{11}$ $M_{\odot}$ in the following redshift ranges: $0.5-1.0$, $1.0-1.5$, $1.5-2.0$, $2.0-2.5$ and $2.5-3.0$. We point out that beyond $z\sim1.5$, the stellar mass range $10^{10.2}-10^{11}$ is populated by less than 20-25 systems. Thus, we limit the use of the MS estimates to the $10^7-10^{10.2}$ $M_{\odot}$ stellar mass range. We estimate the MS by using the best fit relations in stellar mass bins of 0.25 dex to take into account stellar mass uncertainties. We correct the MS to the Kroupa IMF.
    
    \item \cite{2018A&A...615A.146P} use the SED modeling and fitting tool CIGALE to generate flux density priors in the Herschel SPIRE bands in the COSMOS field. These priors are fed into a deblending tool, called $XID+$, to extract flux densities from the SPIRE maps. As the last step, multi-wavelength data are combined with the extracted SPIRE flux densities to constrain SEDs and provide stellar masses and SFRs. These are used to populate the SFR-$M^*$ plane over the redshift range $0.2 < z < 6$. The SED fitting is performed with a Chabrier IMF, BC03 SPS model, and a \cite{2000ApJ...539..718C} extinction law. Star-forming galaxies are selected through the UVJ selection as in \cite{2014ApJ...795..104W}. The MS is estimated in the following stellar mass and redshift ranges: between $10^9$ and $10^{11}$ $M_{\odot}$ at $0.2 < z < 0.5$ and $0.5 < z < 0.8$, between $10^{9.5}$ and $10^{11}$ $M_{\odot}$ at $0.8 < z < 1.1$, $1.1 < z < 1.4$ and $1.4 < z < 1.8$, between $10^{10}$ and $10^{11}$ $M_{\odot}$ at $1.8 < z < 2.3$ and $2.3 < z < 2.9$, between $10^{10,5}$ and $10^{11}$ $M_{\odot}$ at $2.9 < z < 3.9$ and $3.9 < z < 4.9$ and above $10^{11}$ $M_{\odot}$ at $4.9 <z < 6$. We estimate the MS by using the best fit relations in stellar mass bins of 0.25 dex to take into account stellar mass uncertainties. We correct the MS to the Kroupa IMF.
    
    \item \cite{2018MNRAS.477.3014B} provide a measure of the local MS at $z< 0.05$ in the MaNGA sample \citep{2015ApJ...798....7B}. Star-forming galaxies are selected according to the BPT diagram in any spaxel of the MaNGA map. The SFR is estimated through the integrated, extinction-corrected H$\alpha$ emission over the galaxy region, converted via \cite{1998ARA&A..36..189K} law and with a Chabrier IMF. The MS is estimated as the mean SFR in bins of stellar masses in the range $10^9-10^{11.5}$ $M_{\odot}$. We include in our analysis the mean points at the observed stellar masses and redshift and correct to the Kroupa IMF.
    
    \item \cite{2018ApJ...852..107D} estimate the evolution of the specific star formation rate at a fixed mass of $10^{10.3}$ $M_{\odot}$. The SFR is derived by the evolution of the galaxy stellar mass function of active galaxies of \cite{2017A&A...605A..70D}, under the assumption that the stellar mass growth of a galaxy is mainly driven by the integral of the SFR history, minus the quantity of stellar mass returned to the interstellar medium. The GSMF of active galaxies, and thus, the derived SFR evolution is based on the (M$_{NUV}$-M$_R$) $-$ (M$_R$-M$_J$) colors. The SED fitting technique used in this analysis is based on a Chabrier IMF. The MS is estimated at $z\sim$ 2.22, 2.75, 2.25, 3.75, 5, 6 and 7. We correct the MS to the Kroupa IMF.
    
    \item \cite{2018ApJ...853..131L} use the deep CANDELS observations in the GOODS North and South fields to revisit the correlations between stellar mass and star formation rate in galaxies at $1.2< z< 4$. The quantities are estimated via SED fitting with a Chabrier IMF and a BC03 SPS model. SFGs are selected through the UVJ selection. The MS is estimated at stellar masses above $10^9$ $M_{\odot}$ and in the following redshift bins: $1.2-1.5$, $1.5-2$, $2-2.8$ and $2.8-4$. We correct the MS to the Kroupa IMF.

    \item \citet{Iyer2018} use star formation histories (SFHs) reconstructed via the Dense Basis method of \citet{Iyer2017} at 0.5 < z < 6 in the Cosmic Assembly Near-Infrared Deep Extragalactic Legacy Survey GOODS-S field to study the nature and evolution of MS. They use the reconstructed SFHs as trajectories in SFR-stellar mass plane. This allows them to study galaxies at epochs earlier than observed by propagating them backward in time along these trajectories. To generate spectra corresponding to a galaxy with a given basis SFH, they use the Flexible Stellar Population Synthesis (FSPS) mode \citep{Conroy2009, Conroy2010}. They use a Chabrier IMF, a \citet{2000ApJ...533..682C} attenuation law, and IGM absorption according to \citet{Madau1996} prescription. They study the MS at z = 1, 2, 3, 4, 5, 6 using both direct fits to galaxies observed at those epochs and $SFR-M^*$ trajectories of galaxies observed at lower redshifts. In order to exclude quiescent galaxies, they impose a cutoff, excluding systems that are at a distance larger than 0.4 dex from the best fit MS correlation. We correct the MS to the Kroupa IMF.

    \item \cite{2019MNRAS.483.3213P} provide measures of the local MS at $z< 0.085$ and stellar masses $>10^{10}$ $M_{\odot}$ based on several SFR indicators. These include SDSS dust corrected H$\alpha$ emission \citep{2004MNRAS.351.1151B}, WISE 22 $\mu$m emission, dust corrected UV emission \citep{2016ApJS..227....2S} and {\it{Herschel}} PACS and SPIRE emission at 100, 160, 250, 350 and 500 $\mu$m \citep{2016MNRAS.462.3146V}. We include all of them in the analysis. The SFR are all based on a Chabrier IMF. The MS is estimated as the peak of the SFR distribution in several stellar mass bins in the SFR-stellar mass plane. We include in our analysis the mean points at the observed stellar masses and redshift. We correct the MS to the Kroupa IMF.
    
    \item \cite{2019MNRAS.tmp.2263P} measure the evolution of the MS up to z$\sim$2.5 in the stellar mass range $10^{10}-10^{11.5}$ $M_{\odot}$. The SFR is derived by a combination of IR fluxes based on {\it{Spitzer}} MIPS 24 $\mu$m and {\it{Herschel}} PACS and SPIRE data and NUV emission available in the CANDELS and COSMOS fields. The SFR and stellar masses are based on a Chabrier IMF. The MS is estimated as the peak of the SFR distribution in several stellar mass and redshift bins at $0.3 < z < 0.5$, $0.5 < z < 0.8$, $0.8 < z < 1.2$, $1.2 < z < 1.6$, $1.6 < z < 2.2$ and $2.2 < z < 2.5$. This corresponds to the median SFR in a log-normal distribution. We correct the median into the mean and 

    \item \citet{barro2019} use a WFC3 F160W (H-band) selected catalog in the CANDELS/GOODS-N field containing photometry from the UV to far-infrared, photometric redshifts and stellar parameters derived from the analysis of the multi-wavelength data. Stellar masses are estimated with FAST and are consistent with those of the 3D-HST survey \citep[e.g.][]{2014ApJ...795..104W}. They use a ladder of SFR indicators. The SFR ladder consists of three steps that differ on the amount of SFR indicators that are available for each galaxy, namely, UV, mid-IR, and far-IR. The first step is to compute the SFR from the UV luminosity. In this respect, they estimated the total UV luminosity from SED fitting, and use the relation of \citet{1998ARA&A..36..189K}, applying a correction for dust attenuation. The UV attenuation is inferred directly from the best-fit model to the overall SED, which assumes a \citet{2000ApJ...533..682C} attenuation law. For the second and third steps of the ladder, they obtain the total SFR by adding the contributions from the IR emission, either from mid-IR or from far-IR, when available, and the observed, unobscured UV component. They use the calibration of \citet{1998ARA&A..36..189K} corrected to a Chabrier IMF. The authors state that stellar masses and star formation rates are consistent with previous results. The MS is estimated in the redshift bins $0.5 < z < 1.0$, $1.0 < z < 1.8$, and $1.8 < z < 3.0$ over the stellar mass range $\sim 10^{8.6}-10^{11.3}$ $M_{\odot}$. It is measured as the running median of the star-forming galaxies identified with the UVJ criterion. We correct the median to the mean and to the Kroupa IMF. 
    
    \item \cite{2020ApJ...899...58L} measure the evolution of the MS in the COSMOS field by exploiting the VLA-COSMOS 3 GHz Large Project dataset. The MS is estimated in the range $0.3 < z < 6$ in the stellar mass range $10^{9}-10^{11}$ $M_{\odot}$ with different completeness limits as a function of the redshift. The median SFR is estimated via stacking in radio data in several redshift and stellar mass bins. We consider here only the bins where the prior galaxy sample is complete in stellar mass, as indicated by the authors. The SFR is estimated with a Chabrier IMF. SFGs are selected as in \cite{2013A&A...556A..55I} through the NUVRJ color-color selection. We correct the median to the mean and to the Kroupa IMF. 

    \item \citet{Thorne2021} apply {\it{ProSpect}} \citep{Robotham2020} in a parametric mode to multiwavelength photometry from the {\it{Deep Extragalactic VIsible Legacy Survey}}  \citep[DEVILS,][]{Davies2018} in order to measure stellar and dust masses and SFR for galaxies in the COSMOS field in the $0 < z < 9$ redshift range. They use the new DEVILS $UV$ to $FIR$ photometry derived using {\it{profound}} \citep{Robotham2018}. {\it{ProSpect}} uses the \citet{2003MNRAS.344.1000B} stellar libraries and the Chabrier IMF to model the stellar components. To model dust attenuation in galaxies {\it{ProSpect}} uses the \citet{2000ApJ...539..718C} model. {\it{prospect}} utilizes the \citet{Dale2014} templates to model the re-radiation of photons absorbed by dust into the infrared. The authors, point out that, with respect to other SED fitting code, e.g. MAGPHYS\citep{daCunha2008}, based on the same templates, the benefit of {\it{ProSpect}} lies in the fact that it is extremely flexible in how it processes star formation histories and on the fact that it incorporates evolving galaxy metallicities. The author state also that such differences in the SED fitting approach lead to an increase of the stellar mass estimates by 0.2 dex. The comparison with the MAGPHYS stellar masses over the GAMA fields is shown in Section 5 of \citet{Robotham2020}. No discrepancies, instead, are reported with respect to the galaxy SFR. This is also shown in the comparison provided by \citet{Thorne2021} with the MS estimated by \citet{2020ApJ...899...58L} based on radio emission. These are, in turn, in agreement with the $UV+IR$ based MS of \citet{2014ApJ...795..104W} and \citet{2019MNRAS.tmp.2263P} over the same area. Thus, we correct only the stellar masses by 0.2 dex to bring the MS estimates of \citet{Thorne2021} on the same calibration framework of the other publications. We correct also the median to the mean and to the Kroupa IMF. 

    \item \citet{sherman2021} provide the MS of massive galaxies at stellar masses larger than $10^{11}$ $M_{\odot}$ at $1.5 < z < 3.0$ over an area of 17.5deg$^2$.  They use $EAZY-PY$ to perform SED fitting using twelve Flexible Stellar Population Synthesis \citep[FSPS][]{Conroy2009, Conroy2010} templates in the non-negative linear combination. The $EAZY-PY$ FSPS templates are built with a Chabrier IMF, \citet{Kriek2013} dust law, solar metallicity, and star-formation histories including bursty and slowly rising models. 

    Star-forming galaxies are identified by looking at the SFR distribution per stellar mass bin. The authors identify the minimum corresponding to the green valley galaxies to isolate the star-forming systems above this minimum SFR. The MS is estimated as the average SFR of the star-forming galaxies in 4 stellar mass and 4 redshift bins. We correct to the Kroupa IMF.

\begin{figure}
\includegraphics[width=\columnwidth]{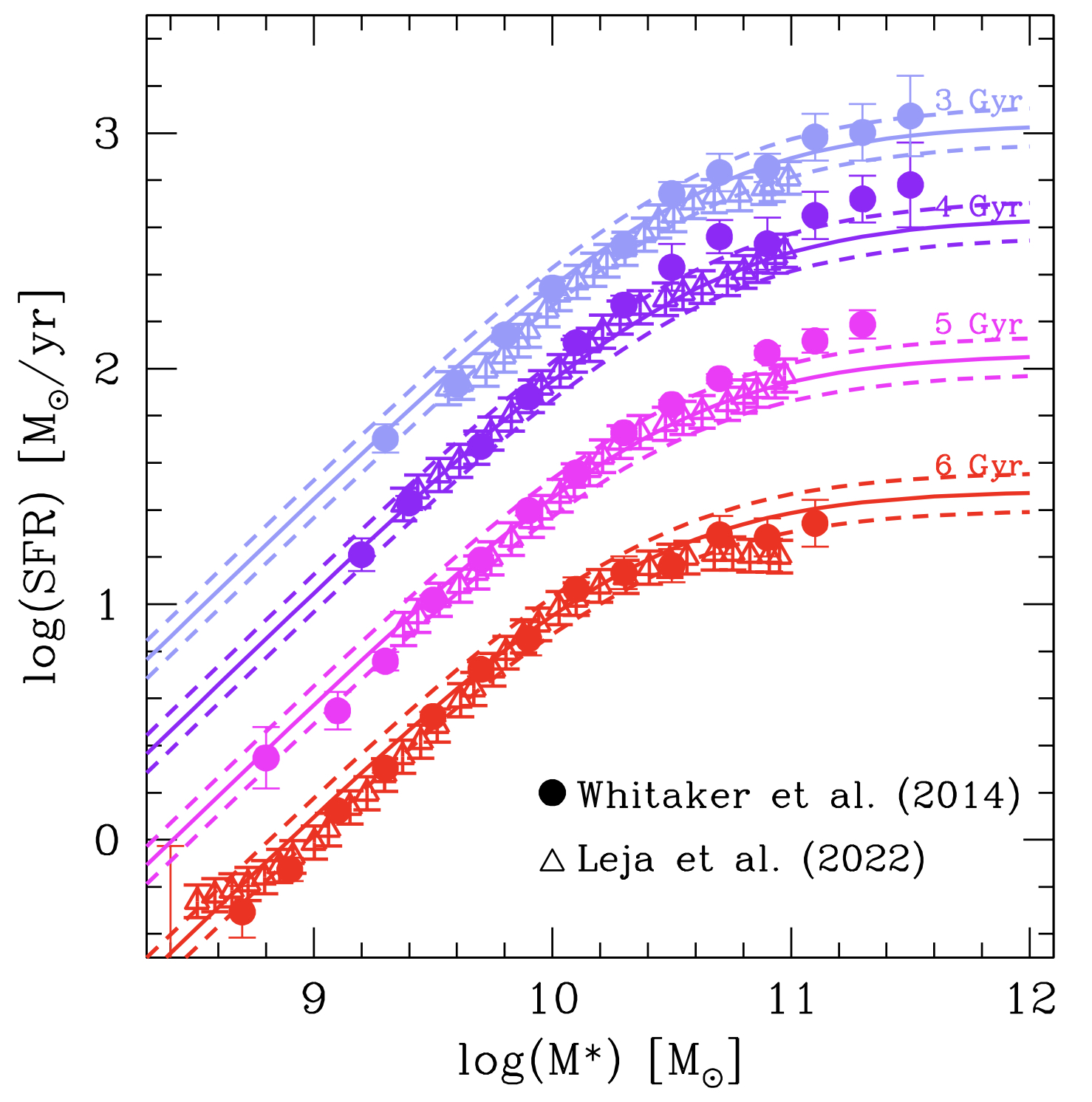}
\caption{Comparison of the MS estimates of \citet{Leja2022} based on the SED fitting code {\it{Proppector}} (empty triangles) and those of \citet{2014ApJ...795..104W} based on the combination of $IR+UV$ in the same deep fields (filled points). The stellar masses of \citet{Leja2022} are corrected by 0.3 dex as indicated in \citet{Leja2019}, for consistency with the stellar masses of \citet{2014ApJ...795..104W}, based on FAST. No correction is applied to the SFR estimates. The MS are color coded as a function of the Universe age as indicated in the figure and are displaced by 0.4 dex from one another for clarity. The solid line indicates the best fit of the MS as a function of stellar mass and time based on Eq. \ref{Lee_final}, while the dashed lines indicate the best fit 1$\sigma$ error.}
\label{leja_whit}
\end{figure}

    \item \citet{Leja2022} use the panchromatic SED-fitting code {\it{Prospector}} to measure the star-forming Main Sequence across $0.2 < z < 3.0$ using the COSMOS-2015 and 3D-HST UV-IR photometric catalogs. In particular, they measure the star-forming sequence using stellar population properties inferred by the Prospector-$\alpha$ model built in the Prospector SED-fitting code. The model has 14 free parameters, consisting of a 7-component non-parametric star formation history using the ‘continuity’ prior which disfavors sharp changes in SFR(t) \citep{Leja2019}, a two-component dust attenuation model with a flexible dust attenuation curve \citep{Noll2009}, free gas-phase and stellar metallicity, and a mid-infrared AGN component with a free normalization and dust optical depth. Prospector includes dust emission powered by energy balance \citep{daCunha2008} with a SED and nebular emission self-consistently powered by the model stellar ionizing continuum \citep{Byler2017}. As shown in \citet{Leja2019}, {\it{Prospector}}, or its incarnation in {\it{Prospector-$\alpha$}}, provides larger stellar masses with respect to previous SED fitting code as CIGALE \citep{Noll2009, Boquien2019,Yang2020}, MAGPHYS \citep{daCunha2008} or FAST \citep{Kriek2009}, largely used in the other publications collected here (see Appendix \ref{A2} for a discussion on the cause of the discrepancy). The authors indicate that such stellar mass increase is on average 0.3 dex with respect to other codes. \citet{Leja2022} also indicate that the SFRs derived by {\it{Prospector}} are lower with respect to the SFRs based on the $IR+UV$ contribution, derived in \citet{2014ApJ...795..104W}. However, as shown in Fig. \ref{leja_whit}, we find that after correcting the stellar masses by 0.3 dex, to this quantity to the same calibration framework of all other publications, we do not observe any significant discrepancy with respect to the SFR estimated with different indicators, e.g. the $IR+UV$ indicator used in \citet{2014ApJ...795..104W}. Indeed, the SFRs of \citet{Leja2022} are consistent with the MS best fit as a function of redshift and stellar mass, within the observed scatter of 0.08 dex. Thus, we do not apply any correction to the SFR if not to for the Kroupa IMF.
    
    They use a flexible neural network known as a normalizing flow to identify the ridge line of the density distribution in the $log(SFR)-log(M^*)$ plane.

\end{itemize}

\begin{figure*}
\includegraphics[width=2.1\columnwidth]{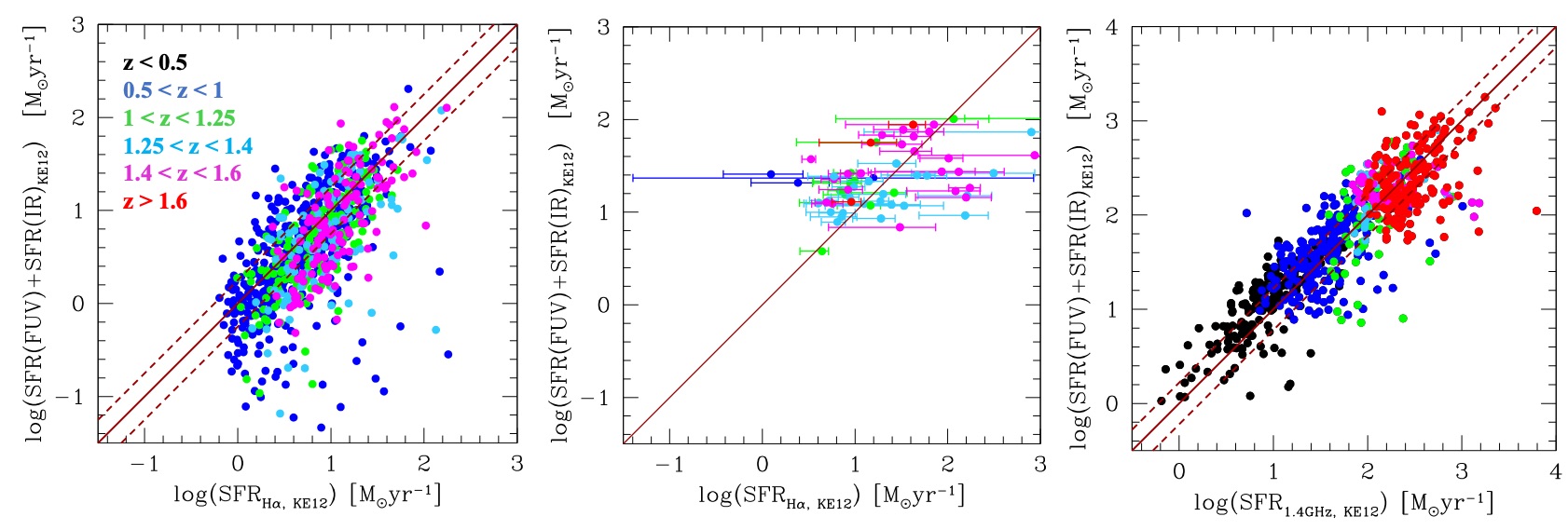}
\caption{{\it{Left panel}}: Comparison of the $UV+IR$ based SFRs versus the SFRs based on H$\alpha$ emission. The H$\alpha$ emission is corrected for dust attenuation on the basis of an average value of the Balmer Decrement, calibrated as a function of the galaxy H$\alpha$ luminosity and stellar mass in \citet{Dominguez2013}. All SFR estimates are obtained according to the KE12 calibration. The points are color-coded as a function of the galaxy redshift. The red solid line shows the 1-to-1 relation and the dashed lines show the dispersion around the relation. {\it{Central panel}}: Same as in the left panel for a subsample of the 3D-HST galaxies where both H$\alpha$ and H$\beta$ exhibit SNR $> 5$ and the Balmer decrement is estimated individually for each system. {\it{Right panel}}: Comparison of the $UV+IR$ based SFRs versus the SFRs based on the radio emission at 1.4 GHz. The color code is the same as in the left panel. The red solid line indicates the 1-to-1 relation and the dashed lines show the dispersion around the relation.}
\label{calibration}
\end{figure*}

\section{The KE12 SFR calibration}
\label{A2}

KE12 offer a collection of different calibrations which leads to consistent SFRs based on different SFR indicators. In this work, we rely mostly on the calibration of \citet{Murphy2011} and \citet{Hao2011} for the $IR$, $UV+IR$, H$\alpha$, and radio 1.4 GHz SFR indicators. \citet{Murphy2011}, in particular, uses the free-free emission measured in the Ka$-$band ($26-40$ GHz) for 10 star-forming regions in the nearby galaxy NGC 6946, including its starbursting nucleus, to compare a number of SFR diagnostics. These diagnostics include non-thermal radio (i.e. 1.4 GHz), total infrared (IR $8-1000$ $\mu$m), and warm dust (i.e., 24 $\mu$m) emission, along with hybrid indicators that attempt to account for obscured and unobscured emission from star-forming regions including $UV + IR$ measurements. By construction, the SFR derived from the indicators calibrated in this way, result to be consistent with each other. Nevertheless, the KE12 calibration is based on the data obtained at $z\sim 0$, and one might ask if this calibration still holds for high redshift galaxies.

There is substantial disagreement between the observed SFRs based on the KE12 or similar calibrations, and the SFRs estimated in simulated galaxies, as provided in state-of-the-art hydro-dynamical simulations. The discrepancy is that the observed SFRs tend to be larger with respect to the predicted ones, in particular, at $0.5 < z < 3$. This questions the reliability of the observed SFRs \citep{Hayward2014,Katsianis2016,Davies2016,Martis2019,Leja2019,Lower2020}. \citet{Katsianis2020}, for instance, argue that the SFR based on the combination of the UV emission and the {\it{Spitzer}} 24 $\mu$m or {\it{Herschel}} 70, 100, 160, 250, 350 $\mu$m emissions are largely overestimated up to 1 dex with respect to the SFRs of simulated galaxies. This work compares the intrinsic SFRs of EAGLE simulated galaxies and those obtained through the KE12 or similar calibrations by analyzing the synthetic SED generated for the same systems by \citet{Camps2018}. Furthermore, \citet{Leja2022} argue that the galaxy star formation histories reconstructed by the SED fitting code {\it{Prospector}} \citep{Leja2017} lead to larger values of stellar masses and lower values of the galaxy SFRs at $0.5 < z < 3.0$,  more in agreement with the SFRs predicted by simulations. This is due to the contribution of an extra component of stars older than 100 Myr, not accounted for in previous SED fitting codes. Indeed, a larger old stellar population would obviously increase a galaxy's stellar mass budget, and it might largely contribute to dust heating, thus, increasing its emission in the mid and far-infrared \citep[see also][]{Viaene2017,Nersesian2019,Leja2019}. \citet{Leja2022} argue that neglecting such a  contribution when calibrating the SFRs, derived from the $IR$ or $IR+UV$ SFR indicators, leads to a clear overestimation of the galaxy SFR as a function of redshift. For this reason, {\it{Prospector}}, which accounts for this contribution, provides lower SFR, bringing in agreement observed and predicted estimates at $0.5 < z < 3$. However, it is worth pointing out that \citet{Leja2022} shows that such contribution is particularly relevant for galaxies below the Main Sequence, as expected \citep{Hayward2014,Nersesian2019}. It is less significant in the MS region, where the larger old stellar population affects mostly the galaxy stellar mass rather than the SFR. This is the main reason why the MS estimated by \citet{Leja2022} are in agreement with previous estimates, based on KE12 calibration, once the stellar masses are corrected for this discrepancy, as shown in Fig. \ref{leja_whit}. 

To further check if the KE12 calibration still holds at high redshift, we use the following approach. If there is an evolution in the fraction of IR emission due to star formation as a function of redshift, as proposed by \citet{Leja2022}, the SFR derived through the $IR$ or $IR+UV$ emissions and other SFR indicators based on KE12, such as the non-thermal radio emission at 1.4 GHz and the nebular H$\alpha$ emission, should no longer be consistent at high redshift. Indeed, the contribution of a larger old stellar population might affect the dust heating and the galaxy IR emission, but it can hardly affect the radio and the H$\alpha$ emissions. The radio emission originating from star-forming galaxies is thought to be caused by Type II and Types Ib supernovae whose remnants are believed to accelerate most of the relativistic electrons in these galaxies. The same supernovae ionize the HII regions as well. Only stars more massive than $\sim 8$ $M_{\odot}$ produce Type II and Ib supernovae, and these have lifetimes of $\sim 3 \times 10^7$ yr, while the relativistic electrons probably live $\sim 100$ yr. Radio observations, therefore, probe very recent star formation in galaxies, and have the advantage that the contribution to radio emission of stellar populations older than 100 Myr is insignificant. Similarly, the H$\alpha$ emission in star-forming galaxies is dominated by nebular emission in the HII regions, ionized by the most massive stars which have a short lifetime \citep{Shivaei2015,Katsianis2016}. The strongest possible contamination is usually due to the AGN emission, rather than old stars. Those might contribute through the emission generated by stellar winds, but this is estimated to be negligible as seen in the local SDSS galaxies. Indeed, it is usually seen in evolved green valley galaxies not classified as star-forming in the BPT diagram \citep{Concas2017}. 

We, thus, compare the SFRs derived from the $IR+UV$ and those derived from the radio 1.4 GHz and H$\alpha$ indicators, calibrated with KE12, in high redshift galaxies. This is to check if they still provide consistent results or if the $IR+UV$ leads to an overestimation at $0.5 < z < 3$. For this exercise, we use UV emission estimated at 2800 \AA\ and the IR emission based on {\it{Spitzer}} 24 $\mu$m data provided in the 3D-HST CANDELS field \citep{2014ApJ...795..104W}. Those are combined to obtain the SFRs according to Eq. \ref{NUV_IR}. The stellar masses are those derived through FAST as described in \citet{Skelton2014}. 

We use the H$\alpha$ emission measured in the public 3D-HST data as described in \citet{Momcheva2016} in the same CANDELS fields. Unfortunately, the SNR of H$\beta$, $[OIII]$ and $[NII]$ emission lines are too low to allow any classification in the BPT diagram \citep{Kewley2013}. Thus, we select all galaxies with an H$\alpha$ emission SNR higher than 5, although we are aware that there might be some contamination by AGN. The H$\alpha$ emission must be corrected for dust attenuation before being converted into SFR through Eq. \ref{Ha}. The low SNR of the H$\beta$ emission prevents to estimate the Balmer decrement if not in a limited subsample of $\sim 80$ galaxies where the SNR is higher than 5 in both H$\alpha$ and H$\beta$ emission lines. Thus, we correct the H$\alpha$ emission with an average estimate of the Balmer decrement as calibrated in \citet{Dominguez2013} as a function of the uncorrected H$\alpha$ luminosity and stellar mass. 

We use the VLA-COSMOS 1.4 and 3GHz catalog \citep{Smolcic2017} to retrieve the galaxy radio emission in the COSMOS field. This catalog provides the galaxy radio luminosity at 1.4 and 3 GHz and isolates a clean subsample of star-forming galaxies with no contamination from AGN identified in the X-rays, infrared, or through the radio excess \citep{Delvecchio2017}. The 1.4 luminosity is converted to SFR through Eq. \ref{radio_sfr}. 

The comparison of the $UV+IR$ based SFRs and the SFRs derived from the radio 1.4 GHZ and H$\alpha$ emissions is shown in Fig. \ref{calibration}. The comparison with the H$\alpha$-based SFRs is shown in the left panel. The points are color-coded as a function of the redshift, as indicated in the figure. We do not observe a clear overestimation of the KE12 $UV+IR$ based SFRs with respect to the H$\alpha$ based SFRs. All data points are distributed around the 1-to-1 relation with a scatter of 0.25 dex, without any clear bias as a function of redshift. On the contrary, we observe outliers with larger values of the H$\alpha$ based SFRs with respect to the $UV+IR$ based SFRs, which we interpret as possible AGN contaminants. The central panel shows the same comparison for the subsample of 3D-HST galaxies where the H$\beta$ emission line SNR allows measuring the Balmer Decrement. Also in this case, we do not observe a clear overestimation of the $ UV+IR$-based SFRs. Nevertheless, we point out that the uncertainties are large also because we do not attempt any correction for absorption and possible line blending issues as done in \citet{Dominguez2013}. The right panel shows the comparison with the SFR based on the radio emission calibrated according to KE12. All data points are distributed around the 1-to-1 relation with a scatter of 0.22 dex and without any clear over- or underestimation as a function of redshift. Few residual sources exhibit a radio excess with respect to the $UV+IR$-based SFRs, which we interpret as radio AGN contaminants not included in the catalog of \citet{Delvecchio2017}.

We, thus, conclude that there is no clear evidence for an overestimation of the $UV+IR$ based SFRs at $0.5 < z < 3$ and that the KE12 calibration allows obtaining consistent SFR estimates up to $z \sim 3$ from the different SFR diagnostics considered here. 

\bsp	
\label{lastpage}
\end{document}